\documentstyle[aps,prb]{revtex}
\input epsf

\newcommand{\subscrpt}[2]{{#1}_{{#2}}}

\newcommand{\bfa}{{\bf a}}

\newcommand{\bfc}{{\bf c}}
\newcommand{\bfcsub}[1]{\subscrpt{\bfc}{#1}}
\newcommand{\bfci}{\bfcsub{i}}

\newcommand{\bfk}{{\bf k}}
\newcommand{\bfn}{{\bf n}}

\newcommand{\bfx}{{\bf x}}
\newcommand{\bfxci}{\bfx + \bfci}
\newcommand{\bfxt}{(\bfx, t)}
\newcommand{\bfy}{{\bf y}}
\newcommand{\bfz}{{\bf z}}

\def\bfsigma{\mbox{\boldmath $\sigma$}}
\def\bfomega{\mbox{\boldmath $\Omega$}}

\newcommand{\bge}{\begin{equation}}
\newcommand{\ee}{\end{equation}}
\newcommand{\bgc}{\begin{center}}
\newcommand{\ec}{\end{center}}
\newcommand{\bgea}{\begin{eqnarray}}
\newcommand{\eea}{\end{eqnarray}}
\newcommand{\bgeas}{\begin{eqnarray*}}
\newcommand{\eeas}{\end{eqnarray*}}

\newcommand{\nisp}{n_i^{S\prime}}

\newcommand{\dt}{\Delta t}
\newcommand{\tpdt}{t+\dt}

\newcommand{\hcc}{H_{\mbox{\scriptsize cc}}}
\newcommand{\hcd}{H_{\mbox{\scriptsize cd}}}
\newcommand{\hdc}{H_{\mbox{\scriptsize dc}}}
\newcommand{\hdd}{H_{\mbox{\scriptsize dd}}}
\newcommand{\smallhalf}{\mbox{\small $\frac{1}{2}$}}

\begin{document}

\title{
\begin{flushleft}
{\footnotesize OUTHP-95-395,
               BU-CCS-9507001,
               comp-gas/9507001}\\[0.2in]
\end{flushleft}
\bf A Lattice-Gas Model of Microemulsions}
\author{
Bruce M. Boghosian,\\
Center for Computational Science, Boston University, \\
3 Cummington Street, Boston,Massachusetts 02215, U.S.A. \\
Peter V. Coveney \\
Schlumberger Cambridge Research,\\
High Cross, Madingley Road, Cambridge CB3 0EL, UK.\\
and \\
Andrew N. Emerton \\
Theoretical Physics, Department of Physics, University of Oxford \\
1 Keble Road, Oxford OX1 3NP, UK\\
[0.3cm]
}
\date{July 10, 1995}
\maketitle

\begin{abstract}
We develop a lattice gas model for the nonequilibrium dynamics of
microemulsions.  Our model is based on the immiscible lattice gas of
Rothman and Keller, which we reformulate using a microscopic,
particulate description so as to permit generalisation to more
complicated interactions, and on the prescription of Chan and Liang for
introducing such interparticle interactions into lattice gas dynamics.
We present the results of simulations to demonstrate that our model
exhibits the correct phenomenology, and we contrast it with both
equilibrium lattice models of microemulsions, and to other lattice gas
models.
\end{abstract}

\vspace{0.2truein}


\section{Introduction}

It is well known that oil and water do not mix, and yet, with the
addition of {\it amphiphile} (or {\it surfactant}) to such systems, one
can observe the formation of a wealth of complex structures. For a
general review see Gelbart {\it et. al.} (1993). The complex structures
themselves arise as a result of the particular physical and chemical
properties of the amphiphile, the most important of these being the
polar nature of the molecules, typically, an ionic head which is
attracted to water, and a hydrocarbon tail which prefers oil.
Consequently there is a strong energy preference for the surfactant
molecules to be absorbed at, and thereby to cause the formation of,
oil-water interfaces.  The aformentioned structures occur in both binary
(surfactant and water or oil) and ternary systems; they are in general
strongly dependent on the relative quantity and nature of the surfactant
molecules present (whether the amphiphile has an ionic or non-ionic
nature, the size of the hydrophobic tail, and so on), and on the
temperature of the system. The equilibrium properties of these complex
structures are conveniently described by the numerous phase diagrams
that have been obtained from experimental investigation (Kahlweit {\it
et. al} 1987). The {\it microemulsion} phase itself occurs only within a
certain region of the ternary phase space. It is of particular
industrial interest because of the very low surface tensions that exist
between the so called ``middle phase'' microemulsion and bulk oil and
water phases ({\it cf.} normal bulk oil-water surface tension), see
Cazabat {\it et. al.} (1982). The complex nature and wide industrial
application of these systems make them ideal subjects for experimental,
theoretical and numerical investigation. Most previous work in this area
has been done on the equilibrium phase behaviour (Gompper \& Schick
1995) of such self-assembling systems; comparatively little research has
been devoted to their nonequilibrium, dynamic properties (Kawakatsu {\it
et. al} 1993). Here we develop a computational technique, with a
theoretical basis, that has the ability to simulate self-assembling
amphiphilic systems under both equilibrium and nonequilibrium
conditions.

Lattice gases have been used as a numerical technique for modelling
hydrodynamics since their introduction in 1986 by Frisch, Hasslacher and
Pomeau (Frisch {\it et al.} 1986) and by Wolfram (Wolfram 1986), who
showed that one could simulate the solution of the incompressible
Navier-Stokes equations using a class of deterministic lattice gases
with discrete Boolean elements. The method was later generalised by
Rothman and Keller (Rothman \& Keller 1988) to enable the simulation of
two immiscible fluids, and we use their model, which has since been
investigated with some degree of rigour (see Sec.~\ref{sec:ilg}), as the
basic starting point for ours.

The simulation of the behaviour of three-phase fluids is a complex and
challenging area of computational science.  Previously, certain
lattice-gas models for three immiscible fluids (Gunstensen \& Rothman
1991a) have been found to exhibit numerical difficulties, including a
very high diffusivity of one fluid into another, very large interfacial
fluctations and limited control over the relative strength of the
surface tension coefficients; such problems have encouraged a trend
towards the study of simpler lattice-Boltzmann techniques (Gunstensen
1992). Furthermore, the complexity of microemulsion behaviour goes well
beyond that of immiscible fluids and so has been out of reach of these
types of models until now.  In this paper we develop a version of such a
model that simulates the nonequilibrium, dynamical properties of
amphiphilic systems, and in particular of microemulsions.  Note that the
use of the basic lattice gas model, in contrast to, for example,
molecular dynamics simulations, gives us the ability to investigate the
important late-time dynamics of these systems (Kawakatsu {\it et al.}
1993) in a computationally tractable manner.  Also, since lattice-gas
models enable the implementation of complex boundary conditions, we can
simulate very realistic interface formation and dynamics and investigate
such systems under flow and within complex media such as porous rock.

The purpose of the present paper is to define and establish the general
validity of our model; we show that our model can reproduce certain
known features of these self-assembling amphiphilic systems.  we show
that our model can reproduce. In Section~\ref{sec:emm} we briefly review
some of the work that has already been done on the modelling of
self-assembling amphiphilic systems.  Sections~\ref{sec:ilg} and
\ref{sec:lgmm} contain the description and formulation of our model,
while the results of simulations we have performed are described in
Section~\ref{sec:ner}. These two-dimensional ($2D$) simulations
demonstrate the ability of our model to capture the phenomenology of
various self-assembling amphiphilic structures in a consistent way,
including the propensity for the surfactant molecules to sit in thin
layers at oil-water interfaces and the arrest of separation of the
immiscible oil-water phases when enough surfactant is present in the
system. In Section~\ref{sec:conc} we draw some conclusions from this
work.

\section{Equilibrium Models of Microemulsions}
\label{sec:emm}

Microemulsion models are characterized by the fundamental length and
energy scales on which the system is described. On this basis we can
categorise the three principal types of theories that have emerged,
namely membrane models, Ginzburg-Landau theories, and microscopic
models.  For a review of such approaches see Gompper and Schick (Gompper
\& Schick 1995) and references contained therein.

Membrane theories take as the basic statistical entities sheets of
amphiphile, whether monolayer or bilayer, thereby providing a universal
description for ternary and binary mixtures. The basic energy scale is
that needed to bend the sheet, and the local free energy per unit area
is written as an expansion in powers of the local curvature. The
membrane approach is most useful for the study of the fluctations of a
single membrane, or of those characterizing a bulk phase whose scaling
behaviour is readily obtained.

Ginzburg-Landau theories are described in terms of order parameters,
which represent microscopic quantities averaged over extended regions of
phase space. The free energy is often simply constructed from symmetry
arguments once the choice of order parameters describing the system has
been made.  The advantage of such theories is that they are simple
enough to permit analytic analysis, to easily describe bulk and
interfacial properties, and to calculate structure functions of the
various phases.  Again the mesoscopic length scales of the microemulsion
must emerge from the theory, as they are much larger than the input
length scales.

The fundamental length scales of microscopic models are molecular ones.
The mesoscopic lengths observed in the experimental analysis of such
systems (scattering experiments and micrographs) must then be derived
{}from the theory. All such theories to date have attempted to derive
generic behaviour and so the distinguishing features of particular
amphiphiles and oils are largely ignored. The interactions included are
invariably of short range, while the amphiphile molecules are treated
either as scalar particles without internal structure or as vectors
whose interaction energies depend on how they align with themselves and
the other entities in the system. An additional simplification that is
often made is to replace the continuum system by a discrete one, which
then reduces the model to the form of a general lattice gas. Advantages
of this lattice approach include its ability to model both bulk phase
and interfacial behaviour. Furthermore, the structure of interfaces,
their tensions and associated wetting properties can be observed at the
molecular level, while the equilibrium structure of the microemulsion
itself can be examined by lattice-based Monte-Carlo simulation
techniques. It is worth noting that hybrid models (Kawakatsu \& Kawasaki
1990) for the simulation of immiscible binary mixtures with surfactant
molecules have been derived; these make use of a combination of
time-dependent Ginzburg-Landau theory for the underlying binary fluid
and discrete particles with dipolar orientation to describe the
surfactant molecules.

Since our model is based on a microscopic treatment of the lattice gas
particles, we look in more detail at some of the various microscopic
models that have been investigated. There are numerous lattice-based
microscopic models, including the three component model of Schick and
Shih (1987), where all three components are found on lattice sites and a
fairly simple Hamiltonian describes the interactions in the system.
This includes a three-particle interaction term that allows the most
obvious property of amphiphiles - their tendency to sit between oil and
water - to be modelled without having to introduce extra degrees of
freedom associated with their directional properties.  This type of
Hamiltonian can be generalized further with a four-particle interaction
term that allows the various structures forming in binary systems,
including amphiphilic bilayers, to be modelled (Gompper \& Schick 1989).

One much studied lattice model, introduced by Widom (Widom 1984; Widom
1986; Widom \& Dawson 1986), again has three species, but in this case
the molecules are confined to the {\it bonds} of a regular lattice.  The
advantage of this model is that it is exactly equivalent to a
spin-$\smallhalf$ Ising model in a magnetic field, with ferromagnetic
nearest-neighbour, antiferromagnetic next-nearest-neighbour, and
three-spin interactions.  The model allows a straightforward
representation of many of the observed microemulsion-like structures,
and it has the correct pattern of phase equilibria, interface phenomena
and interfacial tension (Dawson 1986).  There are many more
generalisations of both the Widom and the three-component models.  One
such example is the Alexander model (Stockfish \& Wheeler 1988), in
which the molecules are not treated symmetrically; oil and water
particles are placed on the lattice sites, at which up to two molecules
can reside, and amphiphiles on the bonds, which are otherwise empty.  As
mentioned above, other microscopic models exist which allow the
amphiphilic molecules to have orientational degrees of freedom.  These
are called vector models and generally consist of a simple Hamiltonian
for a ternary mixture on a lattice, with an additional Hamiltonian that
governs the orientation-dependent interactions. Some of these models
allow the vector amphiphile to have a continuous orientational degree of
freedom (Gompper \& Schick 1989), while others permit the vector
amphiphile to point only towards nearest-neighbour sites (Matsen \&
Sullivan 1990).

In general the phase behaviour and/or phase diagrams obtained from all
these models are reasonable, although for the most part they bear little
actual resemblence to experimentally derived ones.  This is because for
the most part they are given as functions of theoretical interaction
parameters and what one would really like to see is transitions due to
changes in the relative concentration of amphiphile or temperature. It
is intended that our model should have the ability to do this. The
microscopic approach is also particularly appropriate for the study of
self-assembly, in that the amphiphilic molecules are free to form
interfaces or micelles as they choose. However, it is important to note
that, aside from the hybrid model, all the models introduced above have
the ability to model only static, equilibrium phase behaviour.  The
dynamics of such amphiphilic systems are unattainable by these
methods. Another element lacking is any unified approach to binary and
ternary systems, for example, there have been few attempts to describe a
system as it evolves from a balanced ternary one to a binary one as oil,
or water, is withdrawn. The model we present in this paper aims to
correct many of the above mentioned deficiences.

\section{Immiscible Lattice Gases}
\label{sec:ilg}

Lattice gases for the simulation of the flow of immiscible Navier-Stokes
fluids were first introduced by Rothman and Keller in 1988 (Rothman \&
Keller 1988). They generalised the work of FHP to include new degrees of
freedom and collision rules that gave rise to immiscible two-phase flow.
The Rothman-Keller model can be described by supposing that the lattice
gas particles are of two or more different {\it colours}, to denote the
two immiscible species.  They then defined the local colour flux of the
particles leaving a lattice site, and the colour field due to particles
at neighbouring lattice sites, respectively. The local colour flux is
the difference between the two outgoing colour momenta at a site, while
the colour field is defined to be the direction-weighted sum of the
signed colour density at nearest neighbours.  The Rothman-Keller
collision rules were designed such that the ``work'' performed by the
flux against the field is minimized, subject to the constraints of mass
and momentum conservation. This effectively results in sending coloured
particles towards regions of like colour, hence inducing cohesion and
immiscible behaviour. Since then, lattice Boltzmann versions of this
model have been introduced (Gunstensen {\it et. al.} 1991), and much
research has been carried out on the theory (Rothman \& Zaleski 1994),
computer implementation (Gunstensen 1992), and phenomenology (Rothman \&
Zaleski 1994) of these models.  In this section, we show that the
Rothman-Keller model can be derived from an individual particle
description, that will allow us to more easily motivate its
generalisation to include surfactant particles.

We work on a $D$-dimensional lattice, ${\cal L}$, with $n$ lattice
vectors per site.  We denote the lattice vectors by $\bfci$, where
$i\in\{ 1,\ldots,n\}$; we note that rest particles can be accommodated
in this framework by a corresponding zero lattice vector.  The state of
the Rothman-Keller model for $M$ immiscible species at time $t$ is then
completely specified by the quantities $n_i(\bfx,t)\in\{ 0,\ldots,M\}$,
where $i\in\{ 1,\ldots,n\}$ and $\bfx\in {\cal L}$.  We have
$n_i(\bfx,t)=\alpha$ if there is a particle of colour $\alpha\in\{
1,\ldots,M\}$ with velocity $\bfci$ at position $\bfx$ at time $t$, and
$n_i(\bfx,t)=0$ otherwise.  Thus, each site can be in any one of
$(M+1)^n$ different states.  We sometimes find it convenient to use the
alternative representation,
\[
n_i^\alpha(\bfx,t)\equiv\delta_{\alpha,n_i(\bfx,t)},
\]
where $\delta_{\alpha,\beta}$ is the Kronecker delta, though we note
that the $n_i^\alpha(\bfx,t)\in\{ 0,1\}$ are not all independent since
there can be at most one particle of any colour at a particular
position, velocity, and time.

The evolution of the lattice gas for one generation takes place in two
substeps.  In the {\it propagation} substep, the particles simply move
along their corresponding lattice vectors,
\[
n_i(\bfxci,\tpdt)\leftarrow n_i(\bfx,t).
\]
This is followed by the {\it collision} substep, in which the newly
arrived particles change their state in a manner that conserves the mass
of each species,
\bge
\rho^\alpha\bfxt\equiv\sum_i^n n_i^\alpha\bfxt,
\label{eq:rbmass}
\ee
as well as the total $D$-dimensional momentum,
\bge
{\bf p}\bfxt\equiv
   \sum_\alpha^M\sum_i^n\frac{\bfci}{\dt} n_i^\alpha\bfxt,
\label{eq:rbmom}
\ee
where we have assumed for simplicity that the particles carry unit mass.

To further specify the collision process, we partition the $(M+1)^n$
different states of a site into equivalence classes of states that have
the same values for the $M+D$ conserved quantities ($M$ conserved masses
and $D$ conserved components of momentum).  Appendix~\ref{sec:aeq}
contains further information on the derivation of these equivalence
classes. Collisions are thus required to take a state $s$ to another
state $s^\prime$ within the same equivalence class.  If the equivalence
class is of size one, as it would be if there were only one incoming
particle, this uniquely specifies the collision process.  For the more
usual case in which there are many possible outgoing states, we must
specify how to choose a single outcome.

We first consider the case of only two immiscible species ($M=2$); we
denote their colours by $\alpha=B$ or ``blue'' for water, and $\alpha=R$
or ``red'' for oil.  In order to give these two phases cohesion, the
Rothman-Keller model (Rothman \& Keller 1988) favours collision outcomes
that send particles of a given colour to neighbouring sites that are
already dominated by that colour.  To quantify this, we first define the
{\it colour charge} of the particle moving in direction $j$ at position
$\bfx$ at time $t$,
\bge
q_j\bfxt\equiv n_j^R\bfxt-n_j^B\bfxt,
\label{eq:nn1}
\ee
and the total colour charge at a site,
\bgeas
q\bfxt & \equiv & \sum_j^n [n_j^R\bfxt-n_j^B\bfxt] \\
   & = & \rho^R\bfxt - \rho^B\bfxt .
\eeas
We imagine that a colour charge $q$ induces a {\it colour potential},
$\phi (r)= q f(r)$, at a distance $r$ away from it, where $f(r)$ is some
function defining the type and strength of the potential.  Its energy of
interaction with another colour charge $q'$ is then $\hcc=q'\phi
(r)=qq'f(r)$.

Since the collision part of the evolution process of the lattice gas
conserves both the mass of each species and the total $D$-dimensional
momentum, the only contribution to the interaction energy will come from
the propagation phase, where the outgoing colour charges do work in
moving to their new sites. Hence, we consider the interaction energy
between the outgoing particle with colour charge $q_i^\prime\bfxt$ at
$\bfx\in {\cal L}$, and the total colour charge $q(\bfx +\bfy, t)$ at
site $\bfx + \bfy\in {\cal L}$.  If we make an infinitesimal virtual
displacement of the first charge from $\bfx$ to $\bfx +\delta\bfx$ (not
necessarily on a lattice site), the corresponding change in interaction
energy is
\[
\delta \hcc\bfxt =
   q_i^\prime\bfxt q(\bfx + \bfy,t)
   [f(|\bfy - \delta\bfx|) - f(|\bfy|)].
\]
Taylor-expanding in $\delta\bfx$, this becomes
\[
\delta \hcc\bfxt =
   q_i^\prime\bfxt q(\bfx + \bfy,t)
   f_1(y)\bfy\cdot\delta\bfx,
\]
where $y=|\bfy|$, and where we have defined
\bge
f_\ell (y)\equiv\left(-\frac{1}{y}\frac{d}{dy}\right)^\ell f(y)
\label{eq:fdef}
\ee
($\ell$ being a positive integer or zero), giving a concise form for the
derivatives of the function f(y).  Since the outgoing particle with
colour charge $q_i^\prime$ will move in the direction $\bfci$, we let
$\delta\bfx\rightarrow\bfci$.  We then sum over all outgoing colour
charges $i$ at site $\bfx$, and over all sites $\bfy\in {\cal L}$ with
which they might interact to get the total {\it colour work},
\bgeas
\Delta \hcc\bfxt
 &=&
 \sum_i^n \sum_{\bfy\in {\cal L}} q_i^\prime\bfxt
 q(\bfx + \bfy,t) f_1(y)\bfy\cdot\bfci \\
 &=&
 \left(\sum_i^n \frac{\bfci}{\dt} q_i^\prime\bfxt\right)\cdot
 \left(\sum_{\bfy\in {\cal L}} f_1(y)\bfy q(\bfx + \bfy,t)\right)\dt \\
 &=&
 {\bf J}\bfxt\cdot {\bf E}\bfxt\dt,
\eeas
where we have defined the {\it colour flux} of an outgoing state,
\bge
{\bf J}\bfxt\equiv \sum_i^n \frac{\bfci}{\dt} q_i^\prime\bfxt,
\label{eq:cf}
\ee
and the {\it colour field},
\bge
{\bf E}\bfxt\equiv \sum_{\bfy\in {\cal L}} f_1(y)\bfy q(\bfx + \bfy,t).
\label{eq:cff}
\ee
The sum over $\bfy$ extends over the range of the colour interaction,
and will be discussed at length in Sec.~\ref{sec:sten}.  The above
expression for the colour work is identical to that used by Rothman and
Keller if the {\it stencil function} $f_1(y)$ is chosen to select only
nearest neighbour sites.  The stencil function will be discussed in more
detail in Section~\ref{sec:sten}.  For now, we note that we were able to
derive the Rothman-Keller model from the simple assumption that the
particles interact with a potential function $\phi (r)$.

Note that we have intentionally used notation that highlights the
analogy of the Rothman-Keller model to electrostatics.  Specifically,
the colour charge, potential, field and flux are analogous to the
electrostatic charge, potential, field, and the current density,
respectively.  The colour work can then be imagined as a sort of
``resistive heating'' resulting from moving the current of outgoing
colour charges against the colour field in one timestep.

The Rothman-Keller prescription is then to choose the outgoing state
that minimizes $\Delta \hcc$, since this is the one that most
effectively sends particles up the gradient of colour and thereby
induces cohesion.  If multiple outgoing states yield the same minimal
value of $\Delta \hcc$, then the outcome is chosen randomly from among
them.  Rothman and Keller observed that this prescription yields phase
separation when the total particle density on the lattice exceeds a
critical value. Defining the reduced density of a lattice gas as the
average proportion of underlying vectors at an individual lattice site
that will contain a particle of some kind, Rothman and Keller found that
the critical minimum value of reduced density needed for phase
separation was approximately $0.2$.

This prescription was subsequently generalized by Chan \& Liang (1990)
who noted that the colour work, $\Delta \hcc$, can be thought of as a
Hamiltonian function, $H(s,s')$, of the incoming and outgoing states;
and they argued that this Hamiltonian should then be used to construct
Boltzmann weights for choosing the outgoing state.  Specifically, if
${\cal C}(s)$ denotes the equivalence class of states with the same
conserved quantities as state $s$, then one can define a {\it partition
function} for each equivalence class,
\bge
Z(s) = \sum_{s'\in {\cal C}(s)} e^{-\beta H(s,s')}, \label{eq:partf}
\ee
where $1/\beta$ is a temperature-like parameter.  We then define the
collisional state-transition probabilities,
\[
A(s\rightarrow s') =
   \left\{
   \begin{array}{cl}
   \frac{1}{Z(s)}e^{-\beta H(s,s')} & \mbox{if $s'\in {\cal C}(s)$} \\
   0 & \mbox{otherwise}
   \end{array}
   \right.
\]
These transition probabilities are normalized, so that
\[
\sum_{s'} A(s\rightarrow s') = 1,
\]
but they generally do not obey the condition of semi-detailed balance,
\[
\sum_s    A(s\rightarrow s') = 1,
\]
except as $\beta\rightarrow 0$.  Note that this reduces to the
Rothman-Keller model as $\beta\rightarrow\infty$, and to the FHP model
with no interactions as $\beta\rightarrow 0$.  Indeed, Chan and Liang
observed that the phase transition to immiscibility occurs for a
critical value of $\beta$ as well as of density.  More generally, the
Chan-Liang prescription is very useful for constructing lattice particle
simulations that include an interaction Hamiltonian $H$ in a manner
consistent with the conservation of mass and momentum.

\section{A Lattice Gas Model of Microemulsions}
\label{sec:lgmm}

To model microemulsions, we would like to introduce surfactant molecules
in the framework of the Rothman-Keller lattice gas.  To do this, we
first introduce a third species index, say $S$, to represent the
presence or absence of a surfactant molecule.  Thus, since $M=3$, each
site can be in any one of $4^n$ states, and can therefore be
conveniently represented by $2n$ bits.

Surfactant molecules do not contribute to the colour flux and colour
field in the same manner as ordinary coloured particles.  Real
surfactant molecules generally consist of a hydrophilic (often ionic)
portion attached to a hydrophobic (hydrocarbon) portion.  Thus, to
pursue the electrostatic analogy mentioned in Sec.~\ref{sec:ilg},
surfactant particles are best imagined as {\it colour dipoles}.

As in electrostatics, we shall model a colour dipole as a pair of equal
and opposite colour charges, $\pm q$, separated by a fixed displacement,
$\bfa$, in the limit as $\bfa\rightarrow 0$, $q\rightarrow\infty$, and
$q\bfa\rightarrow\bfsigma$ where $\bfsigma$ is the colour dipole vector.
Thus, we characterize the surfactant molecule at position $\bfx$ moving
in direction $i$ by a colour dipole vector $\bfsigma_i\bfxt$.  Note that
the value of $\bfsigma_i\bfxt$ is zero unless $n_i^S\bfxt=1$.  The total
dipolar vector at a site is then denoted by
\bge
\bfsigma\bfxt\equiv\sum_i^n\bfsigma_i\bfxt.
\label{eq:nn2}
\ee
It will be necessary to take scalar products of colour dipole vectors
with other vector and tensor quantities.  Beyond this, however, we leave
their precise representation unspecified for the moment.

\subsection{The Colour / Dipolar Field Interaction}

\begin{figure}
\begin{picture}(100,100)(-170,0)
\put(20,50){\vector(1,0){60}}
\put(80,50){\vector(1,-2){10}}
\put(15,40){$Q$}
\put(10,47){$\bfx$}
\put(50,42){$\bfy$}
\put(80,55){$-q$}
\put(94,25){$+q$}
\put(75,35){$\bfa$}
\end{picture}
\caption{\sl Model of Colour-Dipole Interaction}
\label{fig:de}
\end{figure}
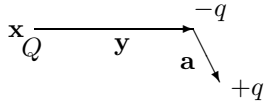

We first consider the work done by a colour charge, $Q$, moving in the
field of a fixed colour dipole, $\bfsigma$, at a displacement $\bfy$, as
shown in Fig.~\ref{fig:de}.
\noindent The energy of the static interaction is
\[
\hcd =
   q Q
   \left[+f(|\bfy+\bfa|) - f(|\bfy|)\right].
\]
In the limit as $\bfa\rightarrow 0$ and $q\rightarrow\infty$, so that
$q\bfa\rightarrow\bfsigma$, this becomes
\[
\hcd = -f_1(y)Q\bfsigma\cdot\bfy.
\]
Since the mass and momentum of particles with colour charge $q\bfxt$ at
$\bfx\in {\cal L}$ are the same before and after the collision part of
the time evolution process, that is colour charge is conserved by the
dynamics, the only local energy change that we need to incorporate into
the Hamiltonian due to the above interaction comes about as a result of
the outgoing particles doing work in moving to their new sites. To
compute this energy change, we note that the interaction energy between
an outgoing particle with colour charge $q_i^\prime\bfxt$ at $\bfx\in
{\cal L}$ and the total dipolar vector $\bfsigma (\bfx+\bfy,t)$ at site
$\bfx+\bfy\in {\cal L}$ is given by
\[
\hcd = -f_1(y)q_i^\prime\bfxt\bfsigma (\bfx+\bfy,t)\cdot\bfy
\]
and that if we make an infinitesimal virtual displacement of the
particle from $\bfx$ to $\bfx+\delta\bfx$ (not necessarily on a lattice
site), the corresponding change in the interaction energy,
$\hcd(\bfy-\delta\bfx) - \hcd(\bfy)$ is
\[
\delta \hcd\bfxt = -q_i^\prime\bfxt
   \bfsigma (\bfx+\bfy,t)\cdot
   \left[
   f_2(y)\bfy\bfy -
   f_1(y) {\bf 1}
   \right]\cdot\delta\bfx,
\]
where $f_\ell (y)$ is defined by Eq.~(\ref{eq:fdef}).  Since the
outgoing particle with colour charge $q_i^\prime\bfxt$ will move in
direction $\bfci$, we let $\delta\bfx\rightarrow\bfci$.  We then sum
over all outgoing colour charges $i$ at site $\bfx$, and over all sites
$\bfy\in {\cal L}$ with which they might interact to get the total work,
$\Delta \hcd$.  We find
\[
\Delta \hcd\bfxt = {\bf J}\bfxt\cdot {\bf P}\bfxt \Delta t,
\]
where we have defined the {\it dipolar field} vector,
\bge
{\bf P}\bfxt\equiv
   -\sum_{\bfy\in {\cal L}}
   \left[
   f_2(y)\bfy\bfy -
   f_1(y) {\bf 1}
   \right]\cdot\bfsigma (\bfx+\bfy,t),
   \label{eq:dfv}
\ee
and where ${\bf 1}$ denotes the rank-two unit tensor.

Thus, we see that the effect of the dipoles at neighbouring sites is to
augment the colour field felt by a colour charge by the amount ${\bf
P}$.  The total work done by the outgoing particles with colour charge
is then
\[
\Delta\hcc + \Delta\hcd = {\bf J}\cdot\left({\bf E}+{\bf P}\right)\dt.
\]
We must now compute the work done by the outgoing dipoles.

\subsection{The Dipole / Colour Field Interaction}

\begin{figure}
\begin{picture}(100,100)(-170,0)
\put(20,50){\vector(1,0){60}}
\put(20,50){\vector(1,2){10}}
\put(15,40){$-q$}
\put(30,75){$+q$}
\put(15,65){$\bfa$}
\put(10,47){$\bfx$}
\put(50,42){$\bfy$}
\put(80,55){$Q$}
\end{picture}
\caption{\sl Model of Dipole-Colour Interaction}
\label{fig:df}
\end{figure}
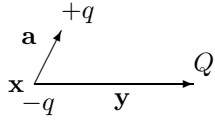

Next, we consider the work done by a colour dipole vector, $\bfsigma$,
moving in the field of a fixed colour charge, $Q$, at a displacement
$\bfy$ as shown in Fig.~\ref{fig:df}.
\noindent The energy of the static interaction is
\[
\hdc =
   q Q
   \left[+f(|\bfy-\bfa|) - f(|\bfy|)\right].
\]
In the limit as $\bfa\rightarrow 0$ and $q\rightarrow\infty$, so that
$q\bfa\rightarrow\bfsigma$, this becomes
\[
\hdc = +f_1(y)Q\bfsigma\cdot\bfy.
\]
In contrast to the case of the outgoing colour charge, the moments of
the colour dipole vectors are not necessarily conserved by the collision
phase of the update, and so the total local energy change which must be
represented here has two parts to it. The first part, at {\it
zeroth}-order, is due to the possibility of the postcollisional
configuration having a different energy from the precollisional
configuration, and the second, at {\it first}-order, is an energy change
resulting from the outgoing colour dipoles doing work in moving to their
new sites. Now, note that the interaction energy between an outgoing
dipole with dipolar vector $\bfsigma_i^\prime\bfxt$ at $\bfx\in {\cal
L}$, and the total colour charge, $q(\bfx+\bfy,t)$, at site
$\bfx+\bfy\in {\cal L}$ is given by
\bge
 \hdc\bfxt= +f_1(y)q(\bfx+\bfy,t)\bfsigma_i^\prime\bfxt\cdot\bfy,
 \label{eq:dczo}
\ee
and that, if we make an infinitesimal virtual displacement of the dipole
{}from $\bfx$ to $\bfx+\delta\bfx$ (not necessarily on a lattice site),
the corresponding change in the interaction energy,
$\hdc(\bfy-\delta\bfx) - \hdc(\bfy)$, is
\bge
\delta \hdc\bfxt = q(\bfx+\bfy,t)
   \left[
   f_2(y)
   \bfsigma_i^\prime\bfxt\cdot\bfy\bfy -
   f_1(y)
   \bfsigma_i^\prime\bfxt
   \right]\cdot\delta\bfx,
   \label{eq:dcfo}
\ee
where $f_\ell (y)$ is defined by Eq.~(\ref{eq:fdef}). The zeroth order
part of the total change in the interaction energy is then obtained
directly from Eq.~(\ref{eq:dczo}) and, letting
$\delta\bfx\rightarrow\bfci$, we can get the first order part from
Eq.~(\ref{eq:dcfo}). Hence the total change in the interaction energy,
$\delta_T\hdc$, is
\[
\delta_T\hdc = f_1(y)q(\bfx+\bfy,t)\bfsigma_i^\prime\bfxt\cdot\bfy +
     q(\bfx+\bfy,t)
        \left[
        f_2(y)
        \bfsigma_i^\prime\bfxt\cdot\bfy\bfy -
        f_1(y)
        \bfsigma_i^\prime\bfxt
        \right]\cdot\bfci.
\]
We then sum over all outgoing colour dipoles $i$ at site $\bfx$, and
over all sites $\bfy\in {\cal L}$ with which they might interact to get
the total {\it dipolar color work}, $\Delta \hdc$. We find, using
Eq.~(\ref{eq:cff}),
\[
\Delta \hdc\bfxt =
 \bfsigma^\prime\bfxt\cdot{\bf E}\bfxt +
 {\cal J}\bfxt : {\cal E}\bfxt \dt,
\]
where the ``double-dot'' notation is short for ${\rm Tr}({\cal J}\cdot
{\cal E})$, and we have defined the {\it total outgoing dipole vector}
\[
\bfsigma^\prime\bfxt\equiv
 \sum^n_i \bfsigma_i^\prime\bfxt,
\]
the {\it dipolar flux tensor}
\bge
{\cal J}\bfxt\equiv
   \sum_i^n\frac{\bfci}{\Delta t}\bfsigma_i^\prime\bfxt,
   \label{eq:df}
\ee
and the {\it colour field gradient tensor}
\bge
{\cal E}\bfxt\equiv
   \sum_{\bfy\in {\cal L}} q(\bfx+\bfy,t)
   \left[
   f_2(y)\bfy\bfy -
   f_1(y) {\bf 1}
   \right],
   \label{eq:cfg}
\ee
and where ${\bf 1}$ denotes the rank-two unit tensor.

Thus, just as the change in colour interaction energy $\Delta \hcc$ due
to an outgoing configuration of colour charges was modelled by Rothman
and Keller as the dot product of a vector flux and a vector field, we
find that the analogous change $\Delta \hdc$ due to an outgoing
configuration of colour {\it dipoles} can be modelled by the {\it
double} dot product of a tensor flux and a tensor field, together with
the addition of a zeroth order term which depends on the colour field
vector itself. To complete the model we need to look at the interaction
energy between two dipoles.

\subsection{The Dipole / Dipole Interaction}

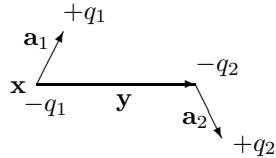
\begin{figure}
\begin{picture}(100,100)(-170,0)
\put(20,50){\vector(1,0){60}}
\put(20,50){\vector(1,2){10}}
\put(80,50){\vector(1,-2){10}}
\put(15,40){$-q_1$}
\put(30,75){$+q_1$}
\put(15,65){$\bfa_1$}
\put(10,47){$\bfx$}
\put(50,42){$\bfy$}
\put(80,55){$-q_2$}
\put(94,25){$+q_2$}
\put(75,35){$\bfa_2$}
\end{picture}
\caption{\sl Model of Dipole-Dipole Interaction}
\label{fig:dd}
\end{figure}

We finally consider the work done by a colour dipole, $\bfsigma_1$,
moving in the field of a fixed colour dipole, $\bfsigma_2$, at a
displacement $\bfy$.  To compute the static interaction energy between
these two dipoles, we return to our fiducial model of a dipole as a pair
of opposite charges, $\pm q$, separated by a fixed displacement, $\bfa$.
Two such dipoles, separated by a distance vector $\bfy$, and
distinguished by subscripts $1$ and $2$, are shown in Fig.~\ref{fig:dd}.
\noindent The static interaction energy between the dipoles is then
\[
\hdd =
   q_1 q_2
   \left[-f(|\bfy-\bfa_1|) - f(|\bfy+\bfa_2|)
         +f(|\bfy+\bfa_2-\bfa_1|) + f(|\bfy|)\right].
\]
In the limit as $\bfa_j\rightarrow\infty$ and $q_j\rightarrow 0$, so
that $q_j\bfa_j\rightarrow\bfsigma_j$, this becomes
\[
\hdd =
   -f_2(y)
   (\bfsigma_1\cdot\bfy)(\bfsigma_2\cdot\bfy) +
   f_1(y)\bfsigma_1\cdot\bfsigma_2.
\]

The analysis of the total interaction energy here is as in the
dipole-colour field case above, since again we have the presence of both
zeroth and first order terms to consider. Similarly, note that the
interaction energy between an outgoing dipole with dipolar vector
$\bfsigma_i^\prime\bfxt$ at $\bfx\in {\cal L}$, and the total dipolar
vector, $\bfsigma (\bfx+\bfy, t)$, at site $\bfx+\bfy\in {\cal L}$ is
given by
\bge
\hdd\bfxt =
   -f_2(y)
   (\bfsigma_i^\prime\bfxt\cdot\bfy)
   (\bfsigma (\bfx+\bfy, t)\cdot\bfy) +
   f_1(y)\bfsigma_i^\prime\bfxt\cdot\bfsigma (\bfx+\bfy, t).
   \label{eq:ddzo}
\ee
and that if we make an infinitesimal virtual displacement of the first
dipole from $\bfx$ to $\bfx+\delta\bfx$ (not necessarily on a lattice
site), the corresponding change in the interaction energy,
$\hdd(\bfy-\delta\bfx)-\hdd(\bfy)$, is
\bgea
\delta \hdd\bfxt & = &
   f_2(y)
   \left\{
   \left[\bfsigma_i^\prime\bfxt \cdot \bfsigma (\bfx+\bfy, t)\right]\bfy +
   \left[\bfsigma (\bfx+\bfy, t) \cdot \bfy\right]\bfsigma_i^\prime\bfxt
   \right. \nonumber \\
   & &
   \left.
   + \left[\bfsigma_i^\prime\bfxt \cdot \bfy\right]\bfsigma (\bfx+\bfy, t)
   \right\} \cdot \delta\bfx -
   f_3(y)
   \left[\bfsigma (\bfx+\bfy, t)\cdot\bfy\right]
   \left[\bfsigma_i^\prime\bfxt\cdot\bfy\right]
   \bfy\cdot\delta\bfx,
   \label{eq:ddfo}
\eea
where the $f_\ell (y)$ are given by Eq.~(\ref{eq:fdef}). Consequently we
can use Eq.~(\ref{eq:ddzo}) to obtain the zeroth order energy change,
while, since the outgoing dipole with dipolar vector
$\bfsigma_i^\prime\bfxt$ will move in direction $\bfci$, letting
$\delta\bfx\rightarrow\bfci$ enables us to obtain the first order part
directly from Eq.~(\ref{eq:ddfo}). The total change in the interaction
energy, $\delta_T\hdd$, is then given by
\bgeas
\delta_T\hdd & = &
   -f_2(y)
   (\bfsigma_i^\prime\bfxt\cdot\bfy)
   (\bfsigma (\bfx+\bfy, t)\cdot\bfy) +
   f_1(y)
   \bfsigma_i^\prime\bfxt\cdot\bfsigma (\bfx+\bfy, t)
   \nonumber \\
   & &
   + f_2(y)
   \left\{
   \left[\bfsigma_i^\prime\bfxt \cdot \bfsigma (\bfx+\bfy, t)\right]\bfy +
   \left[\bfsigma (\bfx+\bfy, t) \cdot \bfy\right]\bfsigma_i^\prime\bfxt
   \right.
   \nonumber \\
   & &
   \left.
   + \left[\bfsigma_i^\prime\bfxt \cdot \bfy\right]\bfsigma (\bfx+\bfy, t)
   \right\} \cdot \delta\bfx -
   f_3(y)
   \left[\bfsigma (\bfx+\bfy, t)\cdot\bfy\right]
   \left[\bfsigma_i^\prime\bfxt\cdot\bfy\right]
   \bfy\cdot\delta\bfx.
   \nonumber
\eeas
We then sum over all outgoing colour dipoles $i$ at site $\bfx$, and
over all sites $\bfy\in {\cal L}$ with which they might interact to get
the total {\it interdipolar colour work}, $\Delta \hdd$.  After some
manipulation, and making use of Eq.~(\ref{eq:dfv}), we find that the
result is
\[
\Delta \hdd\bfxt =
 \bfsigma^\prime\bfxt\cdot{\bf P}\bfxt +
 {\cal J}\bfxt : {\cal P}\bfxt \dt,
\]
where ${\cal J}\bfxt$ is the dipolar flux tensor, and we have defined
the {\it dipolar field gradient} tensor
\bge
{\cal P}\bfxt =
   -\sum_{\bfy\in {\cal L}}
   \bfsigma (\bfx+\bfy, t)\cdot
   \left[
   f_3(y)\bfy\bfy\bfy -
   f_2(y)\bfy\cdot\bfomega
   \right],
   \label{eq:dfg}
\ee
wherein we have in turn defined the completely symmetric and isotropic
fourth-rank tensor,
\[
   \Omega_{ijkl}\equiv
   \delta_{ij}\delta_{kl} +
   \delta_{ik}\delta_{jl} +
   \delta_{il}\delta_{jk}.
\]

\subsection{Expression for the Total Colour Work}

The total interaction work can now be written
\bgeas
\Delta H_{\mbox{\scriptsize int}}
   & = & \Delta \hcc +
         \Delta \hcd +
         \Delta \hdc +
         \Delta \hdd \\
   & = & \left[
           \left(
             {\bf J} +
             \frac{\bfsigma^\prime}{\dt}
           \right)\cdot
           \left(
             {\bf E} + {\bf P}
           \right) +
           {\cal J} :
           \left({\cal E} + {\cal P}\right)
         \right]\dt .
\eeas
The work done to change the kinetic energy is then
\bge
\Delta H_{\mbox{\scriptsize ke}} =
   \sum_\alpha^M \sum_i^n
   \frac{|\bfci|^2}{2}
   \left[
      n_i^{\alpha\prime}\bfxt - n_i^\alpha\bfxt
   \right].
\label{eq:nn8}
\ee
The total work done is then
\bge
\Delta H_{\mbox{\scriptsize total}} =
   \Delta H_{\mbox{\scriptsize int}} + \Delta H_{\mbox{\scriptsize ke}}.
   \label{eq:twork}
\ee
We note that the change in kinetic energy was not included by Rothman
and Keller in their model; neither has it been used in subsequent
studies of their model.  On the other hand, we can think of no good
reason to omit it.  Its presence or absence will not affect the
equilibrium properties of the model, but may impact its nonequilibrium
properties.

\subsection{Stencils}
\label{sec:sten}

We would like to use the total work, $\Delta H_{\mbox{\scriptsize
total}}$ as the Hamiltonian in an implementation of the Chan-Liang
procedure.  To compute this, we will first need the {\it colour flux}
and the {\it dipolar flux} tensor.  These are easily computed directly
{}from their definitions, Eqs.~(\ref{eq:cf}) and (\ref{eq:df})
respectively.  Their computation involves only quantities that are local
to the site $\bfx$.

We will also need the {\it colour field}, the {\it colour field
gradient} tensor, and the {\it dipolar field gradient} tensor.  These
are given by Eqs.~(\ref{eq:cff}), (\ref{eq:cfg}), and (\ref{eq:dfg}),
respectively.  Note that these expressions involve various derivatives
of the potential function $f(y)$, as defined in Eq.~(\ref{eq:fdef}).  In
order to develop a tractable lattice model, however, we shall want to
truncate the range of this function, and so it is unclear how we should
treat its derivatives.  In this section, we shall show how to define a
single function, $g(\bfy)$, which parametrizes the interparticle
potential, and in terms of which we can derive the fields by taking
linear combinations of quantities from some pattern of neighbouring
sites.  These combinations, called {\it stencils}, involve $g(\bfy)$ but
do {\it not} involve its derivatives at all.  We are then free to
truncate $g(\bfy)$ in order to develop a model that involves only
nearest neighbour communication on the grid.

A stencil can be specified as follows: suppose that we have a scalar
field $A (\bfx)$.  Let each site $\bfx$ retrieve the value $A
(\bfx+\bfy)$ from its neighbour at displacement $\bfy$, multiply it by
the coefficient $g(\bfy)$, and sum over all neighbours $\bfy$.  The
result,
\[
   \sum_{\bfy\in {\cal L}} g(\bfy) A(\bfx + \bfy),
\]
is called the {\it scalar stencil} of the scalar field $A$.  We can also
define the {\it vector stencil},
\[
   \sum_{\bfy\in {\cal L}} g(\bfy)\bfy A(\bfx + \bfy),
\]
and the {\it tensor stencil},
\[
   \sum_{\bfy\in {\cal L}} g(\bfy)\bfy\bfy A(\bfx + \bfy),
\]
and so on.  Likewise, given a vector field ${\bf B}(\bfx)$ we can define
its scalar stencil,
\[
   \sum_{\bfy\in {\cal L}} g(\bfy) {\bf B}(\bfx + \bfy),
\]
vector stencil,
\[
   \sum_{\bfy\in {\cal L}} g(\bfy)\bfy {\bf B}(\bfx + \bfy),
\]
and so on.  We require the following isotropy properties of the coefficients
$g(\bfy)$:
\bgea
 & &\sum_{\bfy\in {\cal L}}
    g(\bfy) = G_0
    \nonumber\\
 & &\sum_{\bfy\in {\cal L}}
     g(\bfy)\;\bfy = 0
    \nonumber\\
 & &\sum_{\bfy\in {\cal L}}
     g(\bfy)\;\bfy\bfy = G_2 {\bf 1}
    \nonumber\\
 & &\sum_{\bfy\in {\cal L}}
     g(\bfy)\;\bfy\bfy\bfy = 0
    \nonumber\\
 & &\sum_{\bfy\in {\cal L}}
     g(\bfy)\;\bfy\bfy\bfy\bfy = G_4\bfomega.
    \label{eq:isog}
\eea

As a concrete example, consider a regular Bravais lattice ($|\bfci|=c$),
such that all tensors up to the fourth order formed by the lattice
vectors are isotropic,
\bgea
 & &\sum_j^n 1 = n
    \nonumber\\
 & &\sum_j^n \bfci = 0
    \nonumber\\
 & &\sum_j^n \bfci\bfci = \frac{n}{D} {\bf 1}
    \nonumber\\
 & &\sum_j^n \bfci\bfci\bfci = 0
    \nonumber\\
 & &\sum_j^n \bfci\bfci\bfci\bfci =
    \frac{n}{D(D+2)}\bfomega.
    \label{eq:isoc}
\eea
By choosing the stencil coefficients
\bge
g(\bfy) =
   \left\{
   \begin{array}{ll}
   1 & \mbox{if $|\bfy|=c$} \\
   0 & \mbox{otherwise,}
   \end{array}
   \right.
\label{eq:brav}
\ee
we see at once that we have $G_0=n$, $G_2=n/D$, and $G_4=n/[D(D+2)]$.

We note that lattice gases for fluid applications {\it must} have a
lattice for which Eqs.~(\ref{eq:isoc}) are satisfied in order to ensure
isotropy of the resulting hydrodynamics, so it is always possible to use
Eq.~(\ref{eq:brav}) to define a stencil for such a lattice gas.  It is
possible, however, by judicious choice of the coefficients $g(\bfy)$ to
satisfy isotropy conditions like Eqs.~(\ref{eq:isog}) to higher rank
than those of the lattice vectors, Eqs.~(\ref{eq:isoc}).

\subsection{Stencils for the Colour Field and its Gradient}

We first note that the colour field can be computed directly from its
definition, Eq.~(\ref{eq:cff}), in a straightforward manner.  It is a
vector stencil of the colour charge, with stencil coefficients
$f_1(y)$. We could likewise compute the colour field gradient tensor
directly from its definition, Eq.~(\ref{eq:cfg}), but there is a better
way.  Consider the colour field evaluated at a point $\bfx + \delta\bfx$
(not necessarily at a lattice site).  We have
\[
{\bf E} (\bfx+\delta\bfx,t) =
   \sum_{\bfy\in {\cal L}}
   f_1(|\bfy-\delta\bfx|)
   (\bfy-\delta\bfx)
   q(\bfx + \bfy,t).
\]
Expanding in $\delta\bfx$, this becomes
\bgeas
{\bf E} (\bfx+\delta\bfx,t) &=&
   {\bf E} (\bfx,t) +
   \left\{
      \sum_{\bfy\in {\cal L}}
      q(\bfx+\bfy,t)
      \left[
         f_2(y)\bfy\bfy - f_1(y) {\bf 1}
      \right]
   \right\}
   \cdot\delta\bfx \\
   & & +
   \left\{
      \frac{1}{2}
      \sum_{\bfy\in {\cal L}}
      q(\bfx+\bfy,t)
      \left[
         f_3(y)\bfy\bfy\bfy -
         f_2(y)\bfy\cdot\bfomega
      \right]
   \right\}
   :\delta\bfx\delta\bfx.
\eeas
If we now substitute $\delta\bfx\rightarrow\bfz$ in this equation, and
use it to take the vector stencil of the field, we get
\[
\sum_{\bfz\in {\cal L}} g(\bfz)\bfz{\bf E} (\bfx+\bfz,t) =
   G_2\sum_{\bfy\in {\cal L}} q(\bfx+\bfy)
   \left[f_2(y)\bfy\bfy - f_1(y) {\bf 1}\right],
\]
where we used Eqs.~(\ref{eq:isog}).  Comparing this with
Eq.~(\ref{eq:cfg}), we see that the colour field gradient tensor is
given by the vector stencil of the colour field,
\bge
{\cal E}\bfxt =
   \frac{1}{G_2}
   \sum_{\bfz\in {\cal L}} g(\bfz)\bfz{\bf E} (\bfx+\bfz,t).
\label{eq:cfgtalg}
\ee
In addition to justifying the name of this tensor, this equation is a
much more efficient way to compute it in practice.

\subsection{Stencils for the Dipolar Field and Dipolar Gradient}

Finally, we need to compute the dipolar field and the dipolar field
gradient tensor.  Once again we could compute these directly from their
definitions, Eqs.~(\ref{eq:dfv}) and (\ref{eq:dfg}) respectively, but
there is an easier way.  We first define the scalar field,
\bge
S\bfxt\equiv\sum_{\bfy\in {\cal L}}
   f_1(y)\bfy\cdot\bfsigma(\bfx+\bfy,t),
   \label{eq:sfld}
\ee
using the same stencil coefficients $f_1(y)$ that we used to get the
colour field.  We then consider the value of this field at a point $\bfx
+ \delta\bfx$ (not necessarily at a lattice site).  We have
\[
S(\bfx + \delta\bfx,t) =
   \sum_{\bfy\in {\cal L}}
   f_1(|\bfy-\delta\bfx|)
   (\bfy-\delta\bfx)\cdot\bfsigma(\bfx+\bfy,t).
\]
Expanding in $\delta\bfx$, this becomes
\bgeas
\lefteqn{S(\bfx + \delta\bfx,t) =
   S\bfxt +
   \left\{
      \sum_{\bfy\in {\cal L}}
      \bfsigma (\bfx+\bfy,t)\cdot
      \left[
         f_2(y)\bfy\bfy - f_1(y) {\bf 1}
      \right]
   \right\}
   \cdot\delta\bfx} \\
   & & +
   \left\{
      \frac{1}{2}
      \sum_{\bfy\in {\cal L}}
      \bfsigma (\bfx+\bfy,t)\cdot
      \left[f_3(y)\bfy\bfy\bfy -
            f_2(y)\bfy\cdot\bfomega
      \right]
   \right\}
   :\delta\bfx\delta\bfx.
\eeas
If we now substitute $\delta\bfx\rightarrow\bfz$ in this equation, and
use it to take the scalar, vector, and tensor stencils of the scalar
field $S\bfxt$, we get
\[
\sum_{\bfz\in {\cal L}} g(\bfz) S(\bfx+\bfz,t) =
   G_0 S\bfxt +
   \frac{G_2}{2}
   \sum_{\bfy\in {\cal L}}
   \bfsigma (\bfx+\bfy,t)\cdot\bfy
      \left[
         f_3(y)y^2 - (D+2)f_2(y)
      \right],
\]
\[
\sum_{\bfz\in {\cal L}} g(\bfz)\bfz S(\bfx+\bfz,t) =
   G_2
   \sum_{\bfy\in {\cal L}}
   \bfsigma (\bfx+\bfy,t)\cdot
      \left[
         f_2(y)\bfy\bfy - f_1(y) {\bf 1}
      \right],
\]
and
\bgeas
\sum_{\bfz\in {\cal L}}
   g(\bfz)\bfz\bfz S(\bfx+\bfz,t) &=&
   G_2 S\bfxt {\bf 1}
   +  \frac{G_4}{2} {\bf 1}
   \sum_{\bfy\in {\cal L}}
   \bfsigma (\bfx+\bfy,t)\cdot
      \bfy
      \left[
         f_3(y)y^2 - (D+2)f_2(y)
      \right]\\
   &+&  G_4
   \sum_{\bfy\in {\cal L}}
   \bfsigma (\bfx+\bfy,t)\cdot
      \left[
         f_3(y)\bfy\bfy\bfy -
         f_2(y)\bfy\cdot\bfomega
      \right]
\eeas
respectively, where we used Eqs.~(\ref{eq:isog}).  Comparing these with
Eqs.~(\ref{eq:dfv}) and (\ref{eq:dfg}), we see that we can write
\bge
{\bf P}\bfxt = -\frac{1}{G_2}
   \sum_{\bfz\in {\cal L}}
   g(\bfz)\bfz S(\bfx+\bfz,t),
   \label{eq:bfd}
\ee
and
\bge
{\cal P}\bfxt =
   \left(\frac{G_2}{G_4} - \frac{G_0}{G_2}\right) {\bf 1} S\bfxt
   -\sum_{\bfz\in {\cal L}}
   g(\bfz)\left(\frac{\bfz\bfz}{G_4} - \frac{{\bf 1}}{G_2}\right)
   S(\bfx+\bfz,t),
   \label{eq:cald}
\ee
which are efficient methods to compute the dipolar field vector and the
dipolar field gradient tensor, respectively.

\subsection{Sampling the Outgoing State}
\label{sec:sogs}

We have not yet specified the exact representation of the colour dipole
vectors.  As has been noted, it is necessary that there be a colour
dipole vector $\bfsigma_i\bfxt$ associated with each particle $i$ at
each site $\bfx$ at each time $t$, and that it be possible to take inner
and outer products of this vector with other tensor quantities.  We now
further demand that the {\it magnitude} of each of these dipolar vectors
be a fixed constant $\sigma$ when there is a surfactant particle
present, and be zero when there is not.  That is, we have
\bge
   \bfsigma_i\bfxt = \sigma n_i^S\bfxt \hat{\bfsigma}_i\bfxt,
   \label{eq:sigdef}
\ee
where the dipolar strength $\sigma$ is an input parameter, and where
$\hat{\bfsigma}_i\bfxt$ is a unit vector in the direction of the dipolar
orientation.  This latter quantity can be parametrized by $D-1$
real-valued angles for each direction $i=1,\ldots,n$, for a total of
$n(D-1)$ angles.

We use the Chan-Liang procedure with the Hamiltonian given by the total
work, Eq.~(\ref{eq:twork}).  In this Hamiltonian, we note that the
fields depend only upon the incoming states, $s$, while the kinetic
energy and the fluxes depend upon the outgoing state, $s'$.  More
specifically, the kinetic energy and the colour flux depend only upon
the outgoing occupation numbers, $n_i^{\alpha\prime}\bfxt$, while the
dipolar flux depends as well upon the outgoing dipolar orientations,
$\hat{\bfsigma}_i^\prime\bfxt$.

With this in mind, we adopt the notation $s=(\bfn,\hat{\bfsigma})$,
where $\bfn$ and $\hat{\bfsigma}$ denote the occupation numbers and
dipolar orientations of state $s$, respectively; likewise
$s'=(\bfn',\hat{\bfsigma}')$ denotes the outgoing state.  We can then
write the Hamiltonian in the following form:
\[
H(s,s') = H_0(s, \bfn') +
 \left[
  \frac{\bfsigma^\prime}{\dt}\cdot
  \left[
   {\bf E}(\bfn) + {\bf P}(s)
  \right] +
  {\cal J}(s'):
  \left[
   {\cal E}(\bfn)+{\cal P}(s)
  \right]
 \right]\dt,
\]
where
\bge
H_0 (s, \bfn')\equiv\Delta H_{\mbox{\scriptsize ke}}(\bfn,\bfn') +
   {\bf J}(\bfn')\cdot
   \left[
   {\bf E}(\bfn) + {\bf P}(s)
   \right]\dt,
   \label{eq:hn}
\ee
and where we have suppressed space and time dependences in favour of
state dependences for clarity.  Using Eqs.~(\ref{eq:df}) and
(\ref{eq:sigdef}), we can then write
\[
\left[\frac{\bfsigma^\prime}{\dt}\cdot
 \left[
  {\bf E}(\bfn) + {\bf P}(s)
 \right] +
 {\cal J}(s'):
 \left[
  {\cal E}(\bfn)+{\cal P}(s)
 \right]\right]\dt =
 \sum_i^n\nisp\hat{\bfsigma}_i^\prime\cdot {\bf A}_i(s),
\]
where we have defined the vectors
\bge
   {\bf A}_i(s)\equiv\sigma
   \left[
    \left(
     {\bf E}(\bfn) + {\bf P}(s)
    \right) +
    \left(
     {\cal E}(\bfn)+{\cal P}(s)
    \right)\cdot\bfci\dt
   \right]
   \label{eq:nn3}
\ee
for $i=1,\ldots,n$.

The partition function of Eq.~(\ref{eq:partf}) is then obtained by
summing over all possible outgoing occupation numbers and integrating
over all outgoing dipolar orientations,
\bgeas
Z(s) &=&
   \sum_{\bfn'} e^{-\beta H_0(s,\bfn')}
   \int d\hat{\bfsigma}_1^\prime
   \cdots
   \int d\hat{\bfsigma}_n^\prime
   \exp\left[
   -\beta\sum_i^n\nisp\hat{\bfsigma}_i^\prime\cdot {\bf A}_i(s)
   \right]\\
     &=&
   \sum_{\bfn'} e^{-\beta H_0(s,\bfn')}
   \prod_i^n
   \int d\hat{\bfsigma}_i^\prime
   \exp\left[
   -\beta\nisp\hat{\bfsigma}_i^\prime
   \cdot
   {\bf A}_i(s)
   \right].
\eeas
The probability distribution of outgoing states is then
\[
{\cal P}(\bfn',\hat{\bfsigma}')
   = \frac{1}{Z(s)}
     e^{-\beta H(s,s')}.
\]
The reduced probability distribution of outgoing occupation numbers
(without regard to dipolar orientation) is then
\[
P(\bfn') =
   \frac{1}{Z(s)}
   e^{-\beta H_0(s,\bfn')}
   \prod_i^n
   \int d\hat{\bfsigma}_i^\prime
   \exp\left[
   -\beta\nisp\hat{\bfsigma}_i^\prime
   \cdot
   {\bf A}_i(s)
   \right].
\]
Our strategy shall be to first sample from $P(\bfn')$ to get the
outgoing occupation numbers, $\bfn'$, and to then sample the dipolar
orientations from
\[
Q_i(\hat{\bfsigma}_i^\prime) =
   \frac{\exp\left[-\beta\nisp\hat{\bfsigma}_i^\prime
                   \cdot
                   {\bf A}_i(s)
             \right]}
        {\int d\hat{\bfsigma}_i^\prime
         \exp\left[-\beta\nisp\hat{\bfsigma}_i^\prime
                   \cdot
                   {\bf A}_i(s)
             \right]}
\]
for $i=1,\ldots,n$.

For example, in two dimensions we can parametrize the orientations
$\hat{\bfsigma}_i^\prime$ by the angles $\theta_i^\prime$, so
\[
\hat{\bfsigma}_i^\prime =
   \hat{\bf x}\cos\theta_i^\prime +
   \hat{\bf y}\sin\theta_i^\prime.
\]
Writing ${\bf A}_i$ in polar form,
\[
{\bf A}_i(s) = A_i(s)\left[\hat{\bf x}\cos\phi_i(s) +
                           \hat{\bf y}\sin\phi_i(s)
                     \right],
\]
where
\bge
A_i(s)\equiv\sigma
   \left|
      \left[
         {\cal E}(\bfn)+{\cal P}(s)
      \right]\cdot\bfci + {\bf E}(\bfn) + {\bf P}(s)
   \right|\dt
   \label{eq:nn4}
\ee
and
\bge
\phi_i(s)\equiv\arg\left\{(\hat{\bf x} + i\hat{\bf y})
   \cdot\left(
   \left[{\cal E}(\bfn)+{\cal P}(s)\right]
   \cdot\bfci +{\bf E}(\bfn) + {\bf P}(s)\right)\right\},
   \label{eq:nn5}
\ee
and performing the integration we see that the partition function is
\[
Z(s) = (2\pi)^n
       \sum_{\bfn'} e^{-\beta H_0(s,\bfn')}
       \prod_i^n I_0(\beta\nisp A_i(s)),
\]
where $I_0(z)$ is the modified Bessel function.  The reduced probability
distribution is then
\bge
P(\bfn') =
       \frac{(2\pi)^n}{Z(s)}
       e^{-\beta H_0(s,\bfn')}
       \prod_i^n I_0(\beta\nisp A_i(s)).
   \label{eq:nn6}
\ee
We sample the outgoing occupation numbers $\bfn'$ from this, and then
determine the outgoing dipolar angles by sampling from
\bge
Q_i(\theta_i^\prime) =
   \frac{\exp\left[-\beta\nisp A_i(s) \cos (\theta_i^\prime-\phi_i(s))\right]}
        {2\pi I_0(\beta\nisp A_i(s))}
   \label{eq:nn7}
\ee
for each $i$ from $1$ to $n$ separately. Details of these sampling
procedures are given in Appendix~\ref{sec:asp}. Note that in the absence
of surfactant particles $\bfn'=0$, so the Bessel functions are then all
unity, and the model reduces to Chan and Liang's generalisation of the
Rothman-Keller model (with the addition of the kinetic energy).

\subsection{Summary of the Algorithm}
\label{sec:sota}

We can now write down the full algorithm for our microemulsion lattice
gas model.  Consider, for example, a triangular grid in two dimensions
($D=2$) with up to seven particles per site ($n=7$) corresponding to the
six lattice directions plus a rest particle.  We use Eq.~(\ref{eq:brav})
for $g(\bfy)$, from which it follows that $G_0=7$, $G_2=3$, and
$G_4=3/4$.  (Note that the rest particle does not contribute to $G_2$
and $G_4$.)  The algorithm is then:

\begin{enumerate}
\item For all sites $\bfx$ and lattice vectors $i$, propagate particle
$n_i\bfxt$ and its respective dipolar angle $\theta_i\bfxt$ to site
$\bfxci$.
\item Each site $\bfx$ computes its colour charge $q\bfxt$ according to
Eq.~(\ref{eq:nn1}).
\item Each site retrieves its neighbours' colour charges, and computes its
colour field ${\bf E}\bfxt$ using Eq.~(\ref{eq:cff}).
\item Each site retrieves its neighbours' colour fields, and computes its
colour field gradient tensor ${\cal E}\bfxt$ using
Eq.~(\ref{eq:cfgtalg}).
\item Each site $\bfx$ computes its total dipole vector $\bfsigma\bfxt$
according to Eq.~(\ref{eq:nn2}).
\item Each site retrieves its neighbours' total dipole vectors, and
computes the scalar field $S\bfxt$ according to Eq.~(\ref{eq:sfld}).
\item Each site retrieves its neighbours' scalar fields $S$, and computes
its dipolar field vector ${\bf D}\bfxt$ and dipolar field gradient
${\cal D}\bfxt$ according to Eqs.~(\ref{eq:bfd}) and (\ref{eq:cald}),
respectively.
\item Each site computes its vectors ${\bf A}_i$ according to
Eq.~(\ref{eq:nn3}), and converts them to polar form according to
Eqs.~(\ref{eq:nn4}) and (\ref{eq:nn5}).
\item Each site uses its state $s$ to determine its equivalance class
${\cal C}(s)$, and enumerate the allowed outgoing states $s'$.
\item For each outgoing state $s'$, we compute the colour flux ${\bf
J}(s')$ according to Eq.~(\ref{eq:cf}).
\item For each outgoing state $s'$, we compute the kinetic energy change
$\Delta H_{\mbox{\scriptsize ke}}$ according to Eq.~(\ref{eq:nn8}).
\item For each outgoing state $s'$, we compute the dipole-independent
part of the Hamiltonian, $H_0(s,\bfn')$, according to Eq.~(\ref{eq:hn}).
\item For each outgoing configuration $\bfn'$, we calculate the
probability $P(\bfn')$ according to Eq.~(\ref{eq:nn6}), and normalize
these over all outgoing configurations.
\item The final outgoing state $\bfn'$ is then sampled from $P(\bfn')$.
\item Given $\bfn'$, the final outgoing dipolar angles
$\theta_i^\prime\bfxt$ are sequentially sampled from
$Q_i(\theta_i^\prime)$ given by Eq.~(\ref{eq:nn7}).
\item Go to step 1.
\end{enumerate}

For simulations using our model on moderately sized lattices in $2D$ we
have employed the algorithm as it is written above.  For larger system
sizes, however, it is possible that this algorithm may be prohibitive in
terms of computer memory. In this case simpler versions may be worked
out in which the dipolar vector directions are discrete rather than
continuous, and/or in which a Metropolis Monte Carlo procedure is used
to select the outgoing state. Although the simulations reported in this
paper have been undertaken using a basic $C$ code, the algorithm is
parallelizable, using either the message-passing or data-parallel
paradigms.  We expect to use massively parallel supercomputers to
achieve larger system sizes and later times in future simulations.

\section{Simulations}
\label{sec:ner}

As mentioned earlier there are certain basic properties of
self-assembling amphiphilic systems that it is essential our model be
able to reproduce.  In this section, we describe some of these features
and our efforts to reproduce them with the lattice-gas model defined
above.

\subsection{Common Properties of Microemulsions}

In general, the addition of a small amount of surfactant to a system of
oil and water will not alter the two-phase coexistence; the added
amphiphile will partition itself between the two phases. However, if
there is enough amphiphile present in the system to overcome the
tendency of oil and water to phase separate, then a fluid phase can form
in which the oil and water are solubilized and the surfactant tends to
be ordered in some way. This can result in a finite concentration of
oil-in-water (o/w) or water-in-oil (w/o) droplets, usually called {\it
micelles}, forming within the fluid, or alternatively in the formation
of sheets of amphiphile separating coherent regions of oil and water
(sometimes called {\it lamellar phase}).  If the concentrations of the
oil and water in the system are not very different then both coherent
regions will span the system, and the fluid is said to be {\it
bicontinuous}. Fluids existing in these droplet and bicontinuous phases
are termed microemulsions.  Note that they are characterized by
extensive amounts of internal interface.

Self-assembly also occurs in the two-component binary systems of water
and amphiphile.  The terms micelle and lamellae are also used in the
binary case, when they refer to an often spherical cluster of surfactant
molecules sitting within bulk water (or an inverse micelle if in bulk
oil), and sheets of water separated by amphiphilic bilayers,
respectively.  Typically, at amphiphile concentrations below a critical
value, called the {\em critical micelle concentration}, the amphiphile
molecules exist as isolated {\it monomers} in the bulk water (or oil).
Above the critical micelle concentration, the amphiphile molecules
cluster into micelles of a characteristic average size, consisting of a
well-defined number of monomers; although there are clusters of varying
sizes present, the distribution of sizes is sharply peaked about one
value.  Lamellar phases are found at still higher amphiphile
concentrations.

\subsection{Definition of Coupling Constants}

In order to incorporate the most general form of interaction energy
within our model system, we introduce a set of coupling constants
$\alpha, \mu, \epsilon, \zeta$, in terms of which the total interaction
work can then be written as
\bge
\Delta H_{\mbox{\scriptsize int}}
       =   \alpha \Delta \hcc +
         \mu \Delta \hcd +
         \epsilon \Delta \hdc +
         \zeta \Delta \hdd.
\label{eq:tiw}
\ee
We carry out simulations using the algorithm described in
Section~\ref{sec:sota}; however, we have not yet defined the functions
$f_\ell (y)$ that appear in Eq.~(\ref{eq:fdef}) - in particular we need
to specify the value of $f_1(y)$ since it appears in all four of the
terms in $\Delta H_{\mbox{\scriptsize int}}$. Refering back to
Eq.~(\ref{eq:cff}) we see that if we set
\bge
f_1(y) =
 \left\{
  \begin{array}{ll}
   1 & \mbox{if $y=1$} \\
   0 & \mbox{otherwise}
  \end{array}
 \right.
 \label{eq:stencil}
\ee
then we retrieve exactly the equation used by Rothman and Keller for
their definition of the colour field. In the present paper, we shall
employ this simple prescription and leave any more involved usage of
stencils to a later date.  Note that we also have to specify the
temperature-like parameter $\beta^{-1}$ (Eq.~(\ref{eq:partf})), the
kinetic energy term $\Delta H_{\mbox{\scriptsize ke}}$
(Eq.~(\ref{eq:nn8})), and the dipolar strength $\sigma$
(Eq.~(\ref{eq:sigdef})).  Again, as simplifications for the current
preliminary analysis of the model, we have set $\beta$ and $\sigma$ to
unity and the kinetic energy change $\Delta H_{\mbox{\scriptsize ke}}$
to zero throughout.  Unless otherwise stated, all results described
below have been obtained using $2D$ lattices of grid size $128 \times
128$ with periodic boundary conditions in both dimensions. The colour
pictures showing the time evolution of systems are coded as follows;
black is equivalent to water, green to surfactant and red to oil.

\subsection{Oil-Water System}

\begin{figure}
\begin{center}
\leavevmode
\hbox{
\epsfxsize=6.4in
\epsffile{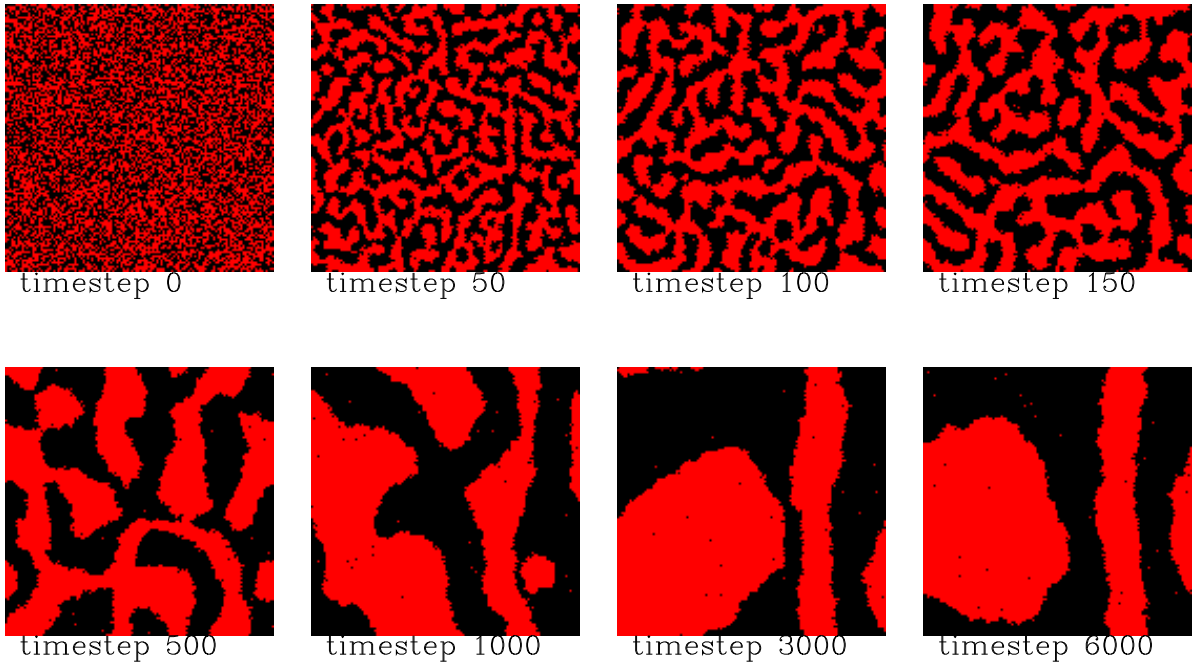}}
\end{center}
\caption{{\sl Nonequilibrium behaviour of a two-immiscible-fluid
lattice gas, consisting of snapshots of the evolving two dimensional system
taken at the timesteps shown.}}
\label{fig:tif}
\end{figure}

We begin by noting that when no surfactant particles are present our
model reduces to Chan and Liang's generalization of the Rothman-Keller
model for two immiscible fluids (Rothman \& Keller 1988).  With no
surfactant particles present in the system, the only term that
contributes numerically to the collision process is $\alpha \Delta
\hcc$.  We set $\alpha=\beta=1.0$ and perform a simulation that has a
random initial configuration with equal amounts of oil and water in the
system. The reduced density (i.e. the proportion of lattice vectors at
each site that contain a particle) is $0.55$.  In this parameter regime,
the Chan-Liang model does exhibit phase separation with positive surface
tension, as is evident from Fig.~\ref{fig:tif}, which illustrates the
nonequilibrium behaviour of the immiscible lattice gas.  In displaying
the results graphically, we have used a majority rule to define the
displayed colour of each site; if the numbers of oil and water particles
at a site are equal then the displayed colour is selected randomly with
probability $\smallhalf$. The behaviour shown in the figure is, indeed,
effectively the same as that obtained by Rothman and Keller. If left to
run for a large enough time the system would eventually reach the
completely separated state of two distinct layers of fluid.

\begin{figure}
\begin{center}
\leavevmode
\hbox{
\epsfxsize=6.4in
\epsffile{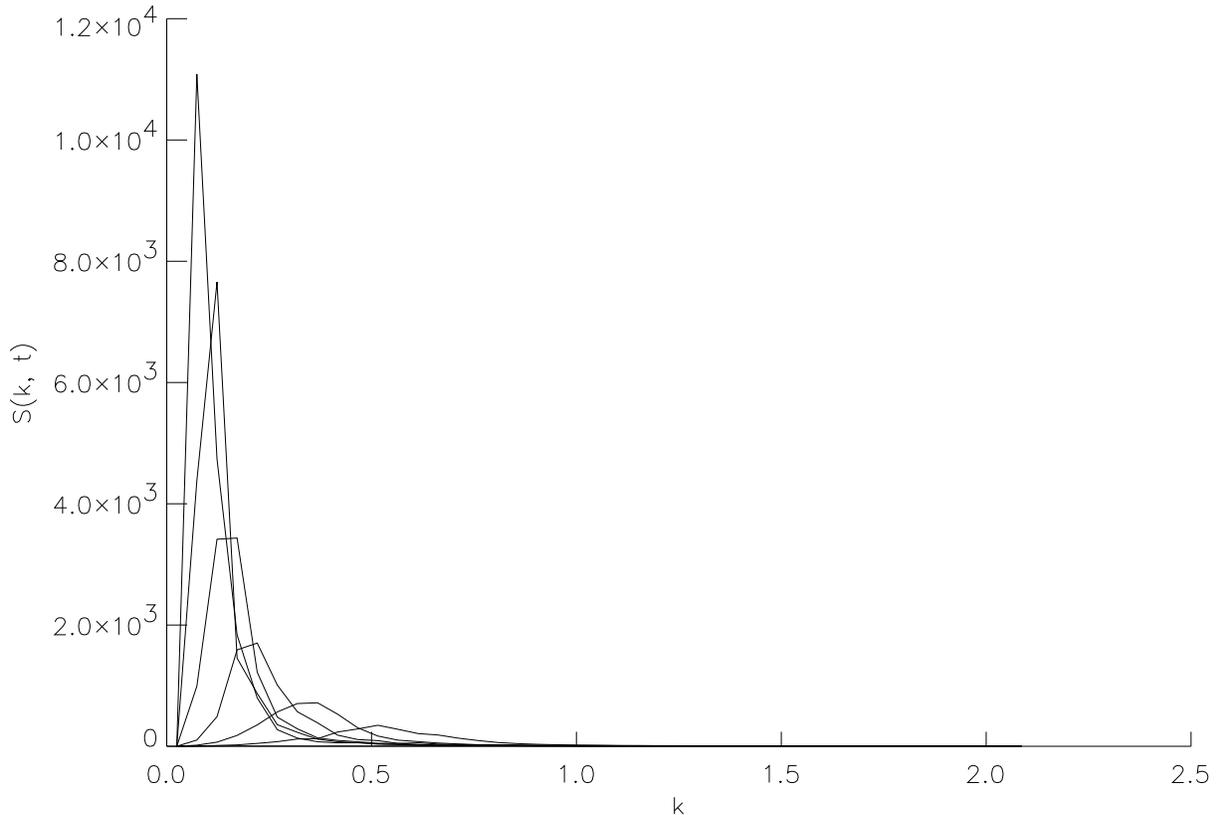}}
\end{center}
\caption{\sl Temporal evolution of $S(k, t)$. Timesteps shown  are, from
bottom to top, t = 0, 80, 200, 400, 800, 1400, 2000.}
\label{fig:sftif}
\end{figure}

To make a detailed comparison between the immiscible oil-water fluid
behaviour shown here and later simulations in which we introduce
surfactant to the system, we make use of {\it circularly averaged
structure functions}. We first calculate (Roland \& Grant 1989) the
two-dimensional structure factor of the colour charge, $s(\bfk, t)$,
\[
s(\bfk, t) = \left<\frac{1}{N}\left|\sum_{\bfx} (q(\bfx) - q^{av})
                   e^{i\bfk\cdot\bfx}\right|^2\right>,
\]
where $\bfk = \left(2 \pi / L \right) \left( m {\bf i} + n {\bf j}
\right)$, $m,n = 1,2,...,L$, q(\bfx) is the total colour charge at a
site, $q^{av}$ is the average value of the colour charge, $L$ is the
length of the system and $N = L^2$ is the number of sites in the system.
The circularly averaged structure factor, which is what we actually
evaluate numerically, is given by
\bge
S(k, t) = \frac{\sum_{\hat k}s(\bfk, t)}{\sum_{\hat k} 1},
\label{eq:castf}
\ee
where $k = 2\pi n / L, n = 0,1,2,...,L$, and the sum $\sum_{\hat k}$ is
over a spherical shell defined by $(n - \frac{1}{2}) \le
\frac{|\bfk|L}{2\pi} < (n + \frac{1}{2})$. Note that the resolution of
$S(k, t)$ depends on $k_c$, the cutoff frequency associated with the
lattice, that is, for a real-space sampling interval of $\Delta$ the
cut-off frequency is $1/2\Delta$; above this value of the frequency
there is only spurious information carried as a result of aliasing. In
our case, $k_c = (2\pi/L)n_c$, where we have chosen $n_c$ to be the
maximum possible value, which is half the lattice
size. Figure~\ref{fig:sftif} contains the temporal evolution of $S(k,
t)$ for the case of two immiscible fluids. The data is averaged over
five independent simulations. As time increases, the peak of $S(k, t)$
shifts to smaller wave numbers and its peak height increases.  This
behaviour is indicative of coarsening and will be used as a comparison
for the surfactant-based systems described below.

\begin{figure}
\begin{center}
\leavevmode
\hbox{
\epsfxsize=4.5in
\epsffile{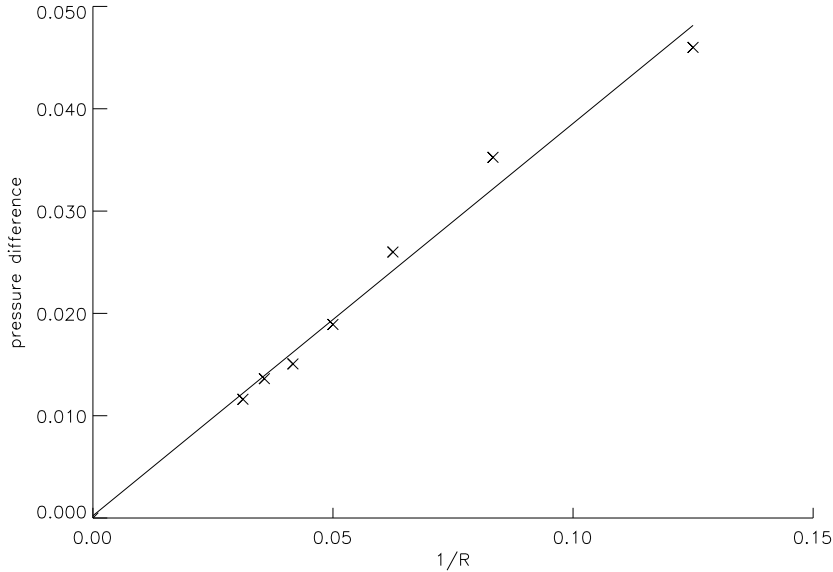}}
\end{center}
\caption{\sl Verification of Laplace's law and estimation of surface
tension}
\label{fig:stb}
\end{figure}

With sufficient averaging and a rescaling of variables it can be shown
(Frisch {\it et. al.} 1986) that the Navier-Stokes equations for an
incompressible fluid are obeyed within homogeneous regions of each of
the fluids.  (There are known problems with galilean invariance due to a
spurious factor in front of the inertial term.  This issue is discussed
later in this paper.)  It is also possible to show, as a result of the
formation of interfaces, that physically realistic interfacial tensions
exist.  In terms of the basic two-species immiscible lattice-gas,
surface tension has been studied from both a theoretical and a numerical
viewpoint in a recent paper by Adler, d'Humi\`{e}res and Rothman (Adler
{\it et. al.} 1994). Using a bubble experiment as described in their
paper, we can investigate the validity of our basic model by evaluating
the surface tension in the immiscible fluid case, noting that the
measurement of surface tension requires the presence of an interface
between two {\it macroscopically} defined phases. We use Laplace's law,
which in two dimensions is
\[
P_{\mbox{in}} - P_{\mbox{out}} = \frac{\sigma}{R},
\]
where $R$ is the radius of the bubble, $P_{\mbox{in}}$ is the average
pressure within the bubble and $P_{\mbox{out}}$ the average pressure
outside. The results from simulations are shown in Fig.~\ref{fig:stb}.
They give good agreement with Laplace's law, and a best-fitting line
through the origin results in an estimate of $\sigma\approx 0.378$.
This is close to the results given by Adler and co-workers, although we
have done significantly less averaging than in their cited paper.

\subsection{Oil-Water-Surfactant Systems}

Next we seek microemulsion-like behavior in ternary mixtures. We note
that the introduction of the surfactant will cause all four of the terms
in Eq.~(\ref{eq:tiw}) to contribute in some way to the interaction
energies of the collision process.  In order that these terms enable us
to reproduce basic microemulsion behaviour, we consider briefly what the
physical contributions of the dipoles ought to be:
\begin{enumerate}
\item In real microemulsions, the surfactant molecules strongly prefer
to sit with their hydrophilic heads in water and their hydrophobic tails
in oil. Since the $\epsilon \Delta \hdc$ term in Eq.~(\ref{eq:tiw})
arises as a result of the interaction of outgoing dipoles with the
surrounding colour field, it is evident that we will want to include a
large contribution from this term as it will favour vector dipoles
aligning across, and sitting as near as possible to, oil-water
interfaces, across which the colour field gradients are at their
greatest.
\item The term $\mu \Delta \hcd$ in Eq.~(\ref{eq:tiw}) results from the
effect of surrounding dipoles on outgoing coloured particles. Since this
will encourage the bending of dipoles around a central colour charge it
may be important when it comes to looking at o/w and w/o microemulsion
droplet phases, but a careful balance will have to be struck between
this and the $\epsilon \Delta \hdc$ term to stop it from destroying the
overwhelming tendency for surfactant to sit at oil-water interfaces,
especially when we are looking to observe bicontinuous structures.
\item The final term $\zeta \Delta \hdd$ in Eq.~(\ref{eq:tiw}) arises as
a result of the interactions between dipoles, and will allow attraction
or repulsion to take place depending on the sign of $\zeta$.  This term
may be of limited use for modelling more general amphiphile-containing
systems because, at present, our model does not differentiate between
the relative strengths of the hydrophobic and hydrophilic interactions
in amphiphile molecules.
\end{enumerate}

After this assessment of the relative values of these constants,
including
\begin{itemize}
\item the need for $\epsilon$ to be relatively large when compared with
      the other constants (see point $1.$ above),
\item the realisation that an effective useful maximum
      value for $\epsilon$ exists, due to constraints imposed
      by the calculation of modified Bessel functions in the
      collision  update process (see Section~\ref{sec:sogs}),
\item the need to compare our results with the known basic lattice-gas
      behaviour, suggesting that initially we maintain the value of
      $\alpha$ used in the two-immiscible-fluid case,
\item some simulation work,
\end{itemize}
we arrived at the following set of coupling constants, that should allow
us to model the various experimentally observed microemulsion
characteristics :
\bge
\alpha = 1.0,
\mu = 0.05,
\epsilon = 8.0,
\zeta = 0.5.
\label{eq:doc}
\ee
We use this as a `canonical' set of coupling constants in the remainder
of this paper.

\subsection{Binary phases: From monomers to micelles}

If one adds a small amount of amphiphile to water, then those surfactant
molecules, while being highly dynamic within the bulk water phase, will
remain distinct from one another and exist as {\it monomers}. Gradually
increasing the amount of amphiphile in the system just increases the
density of these monomers until, at the critical micelle concentration,
the monomers begin to form micelles. These give the system
characteristic structure and should be discernible in our
simulations. The micelles themselves are dynamic objects and are not
necessarily very long lived, since individual molecules are free to
detach themselves, meet with other monomers and/or micelles and rejoin
to form new structures; the kinetics of simple micelle formation can be
modelled on the basis of a Becker-D\"{o}ring theory (Coveney \& Wattis
1995). Individual micelles do not grow without limit; if more surfactant
is added to the system it will form new micelles rather than increasing
the size of those already present.  This is because, energetically, only
a certain number of amphiphile molecules are able to fit around one
central point to produce a micelle structure (for real micelles, the
characteristic number of monomers contained within a micelle depends on
the details of the surfactant's molecular structure).  Consequently we
do not expect to see evidence of micelles coalescing and growing in an
unbounded manner in our simulations.

\begin{figure}
\begin{center}
\leavevmode
\hbox{
\epsfxsize=6.4in
\epsffile{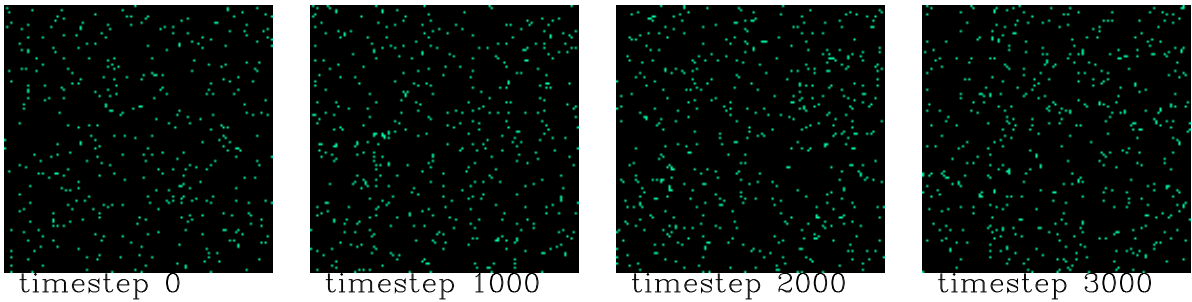}}
\end{center}
\caption{\sl Time evolution of binary water - surfactant phase, for
the case with surfactant - water ratio $1$:$8$.}
\label{fig:bi1}
\end{figure}
\begin{figure}
\begin{center}
\leavevmode
\hbox{
\epsfxsize=6.4in
\epsffile{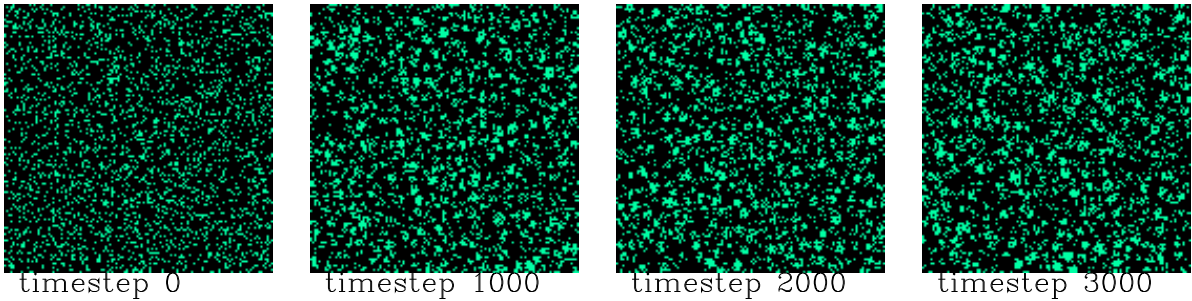}}
\end{center}
\caption{\sl Time evolution of binary water - surfactant phase, for
the case with surfactant - water ratio $1$:$2$.}
\label{fig:bi2}
\end{figure}

We performed two simulations, one with a low concentration of surfactant
and the other with a concentration exceeding the critical micelle
concentration of the system. In both cases, the initial condition
consists of placing water and surfactant particles on the lattice
randomly.  The visual results of the simulations are shown below.  Due
to the limitations of the visualization technique employed, we require
further proof of the existence of structure and so perform a more
quantitative analysis by calculating circularly averaged structure
functions of the surfactant density (Kawakatsu {\it et. al} (1993).  For
consistency, the coupling constants used in both simulations are as
defined in Eq.~(\ref{eq:doc}).  Figure~\ref{fig:bi1} shows the result of
a system containing a surfactant-to-water ratio of $1$:$8$, equivalent
to a initial reduced density of $0.4$ for water and $0.05$ for
surfactant. As before a majority rule is employed to display the type of
particle present at each site.

\begin{figure}
\begin{center}
\leavevmode
\hbox{
\epsfxsize=6.4in
\epsffile{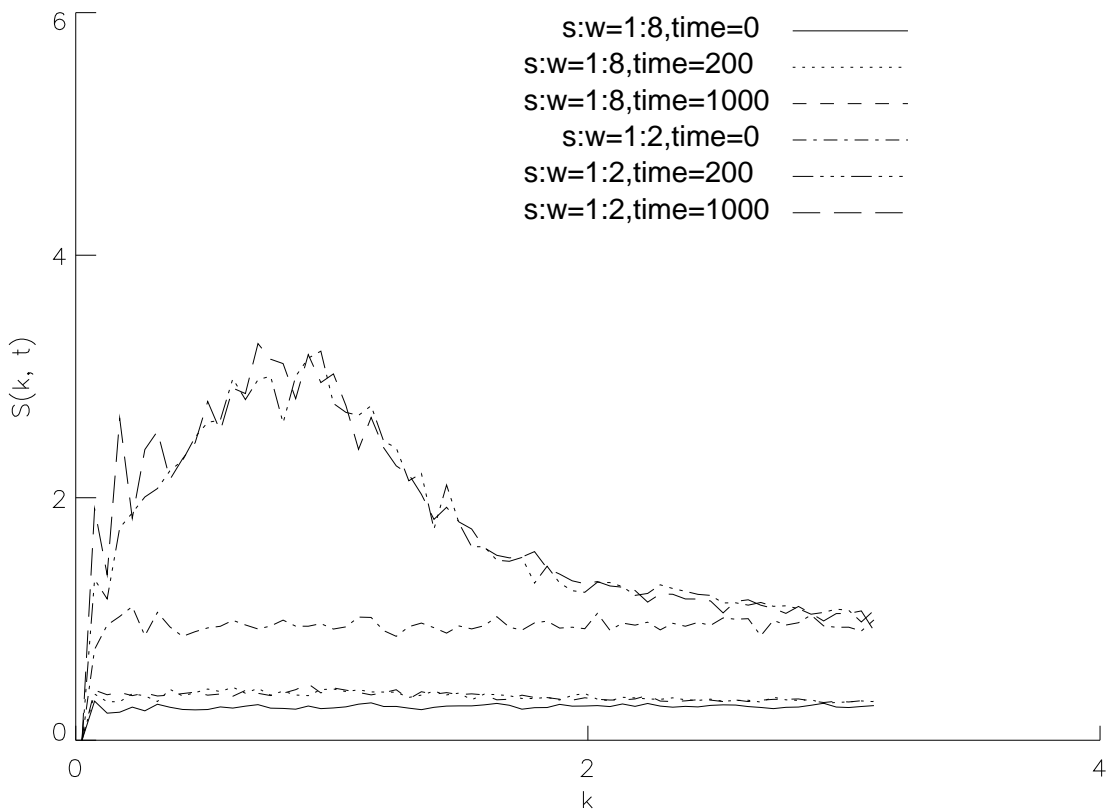}}
\end{center}
\caption{\sl Temporal evolution of surfactant density $S(k, t)$ for binary
water and surfactant mixtures.}
\label{fig:binsf}
\end{figure}

As can be seen in Fig.~\ref{fig:bi1}, with low surfactant concentration
there is very little aggregation of the monomers throughout the
simulation.  This is backed up by the structure factor calculations,
shown in Fig.~\ref{fig:binsf}, which indicate no structure formation
during the timescale of the run.  In stark contrast, Fig.~\ref{fig:bi2},
which has a surfactant-water ratio of $1$:$2$, initial reduced density
$0.4$ for water and $0.2$ surfactant, clearly indicates the formation of
small, structured objects, along with the presence of some monomers.
These structured objects are indeed micelles; they appear in the early
stages of the simulation and, although being highly dynamic, do seem to
maintain their size and shape.  This analysis is confirmed by the
circularly averaged structure factors shown in Fig.~\ref{fig:binsf},
where an average taken over ten independent runs is displayed.  The
graph clearly indicates the formation of structure when the
surfactant-to-water ratio is $1$:$2$, and an absence of structure when
the ratio is $1$:$8$.

\begin{figure}
\begin{center}
\leavevmode
\hbox{
\epsfxsize=6.4in
\epsffile{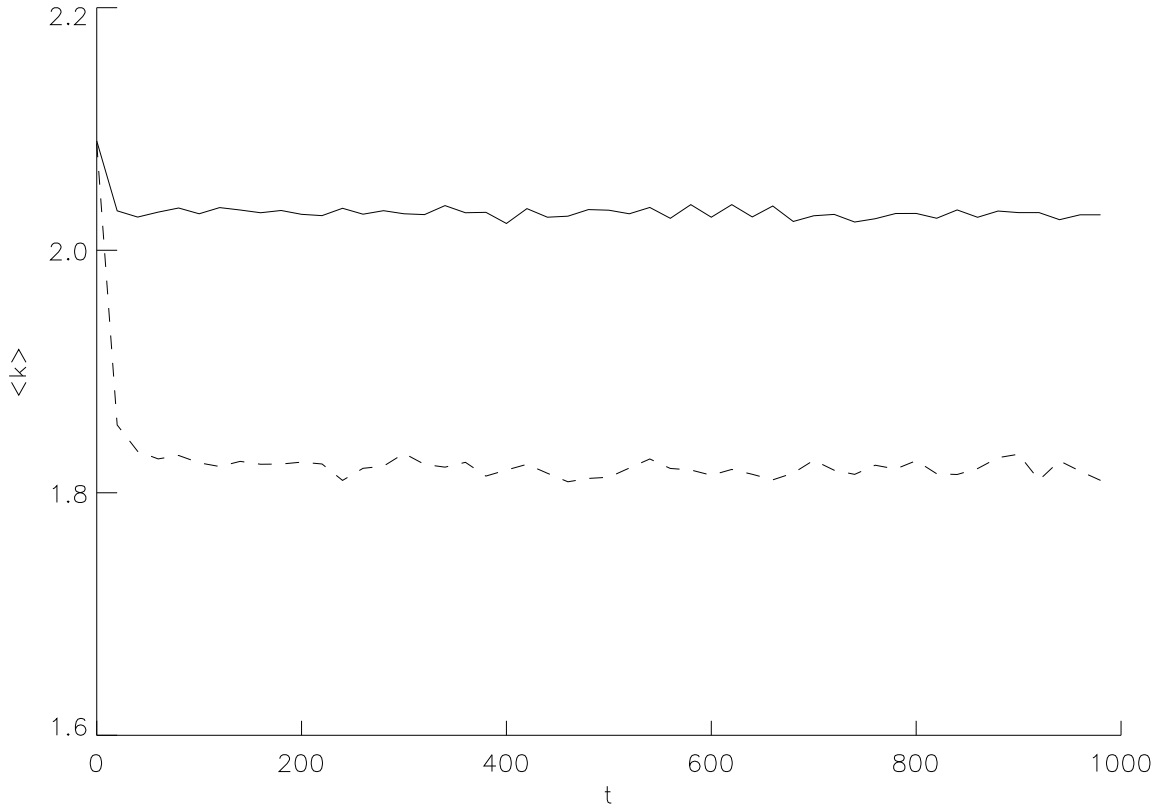}}
\end{center}
\caption{\sl Characteristic wave number $\langle$k(t)$\rangle$ against time
t. The unbroken line on the graph corresponds to the case with surfactant -
water ratio $1$:$8$ and the broken line to the ratio $1$:$2$.}
\label{fig:bing1}
\end{figure}

The fact that the plots at times 200 and 1000 are virtually identical
suggests that the characteristic sizes of the micelle-like structures
saturate at early times during the simulation and that they do not
change dramatically thereafter.  Thus, as expected, the structures do
not grow without limit.

This is made still more clear by Fig.~\ref{fig:bing1}, in which we plot
the temporal evolution of the characteristic wave number,
\[
\langle k(t) \rangle =
 \frac{\sum_{k=0}^{k_c} k S(k, t)}
      {\sum_{k=0}^{k_c}   S(k, t)},
\]
the inverse of which is a measure of the average domain
size (Kawakatsu {\it et. al} 1993).

We see that, in contrast to the case of low amphiphile concentration, we
get initial growth of the surfactant domains which very rapidly levels
off to some constant size.

\subsection{Ternary phases: lamellae}

We first investigate the stability of a lamellar structure, which is
composed of alternating layers of oil-rich and water-rich phases
separated by surfactant molecules. We look at the system with and
without surfactant present in order for a critical comparison to be
made.  In a similar way to Kawakatsu \& Kawasaki (1990) we set up the
initial configuration of the system, resulting in layers of oil and
water eight sites wide, all sites having a reduced density of $0.5$.  It
is clear that if our model is exhibiting the correct behaviour, then we
would expect there to be a critical density of surfactant required at
the oil-water interfaces in order for the layered structure to be
stable. Consequently, we set up a simulation where there is a layer of
surfactant at each of the oil-water interfaces that is just a single
site wide, but with a reduced density on these amphiphilic sites equal
to $0.8$. The results of the simulations undertaken are shown in
Fig.~\ref{fig:lms1}, Fig.~\ref{fig:lms2} and Fig.~\ref{fig:lms3} below.
Figure~\ref{fig:lms1} is the pure oil-water case with $\alpha = 1.0$,
ensuring that the oil and water particles will want to act as immiscible
fluids and so we expect to see phase separation evolving from the
layered initial condition.  Figure~\ref{fig:lms2} has surfactant present
as described above but with coefficients $\alpha = 1.0, \mu = 0.0,
\epsilon = 0.0, \zeta = 0.0$, while Fig.~\ref{fig:lms3} has a similar
amount of surfactant in the system but in this case the coefficients are
as defined by Eq.~(\ref{eq:doc}), and so the full set of interaction
terms in our model are now included. Note that we let the last two
simulations evolve to late times in order to check that we are not just
observing metastable states with long equilibration times which might
arise as a result of the particular set of initial conditions chosen.

\begin{figure}
\begin{center}
\leavevmode
\hbox{
\epsfxsize=6.4in
\epsffile{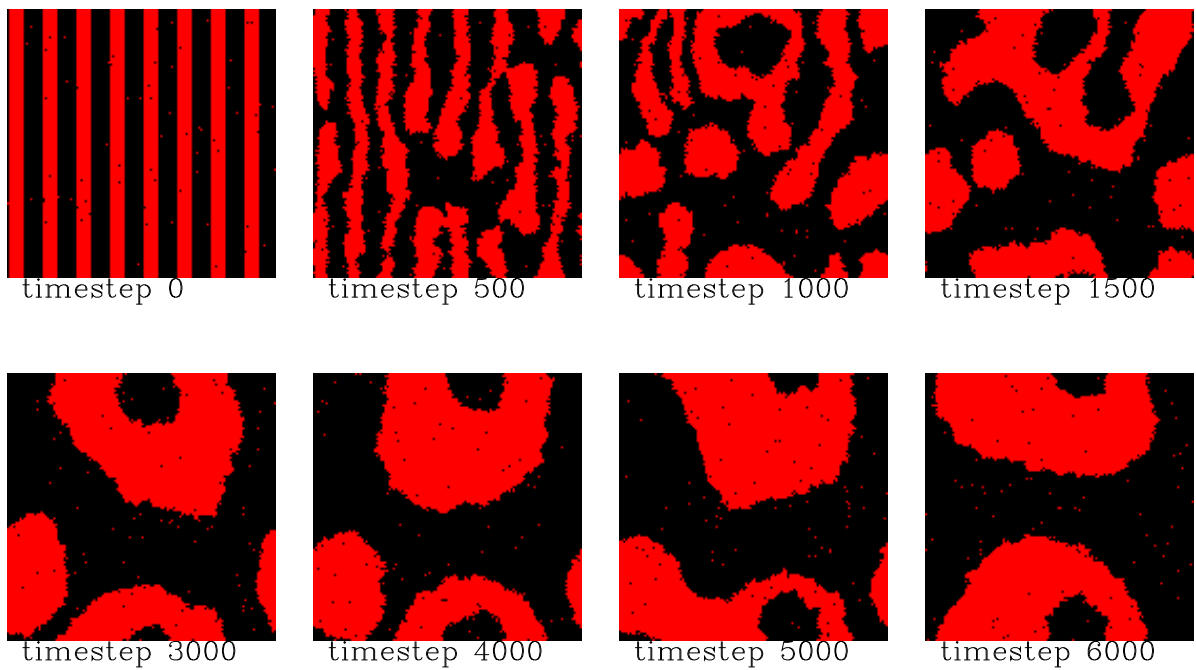}}
\end{center}
\caption{\sl Time evolution from initial lamellar configuration, for
the case without surfactant.}
\label{fig:lms1}
\end{figure}

The probabilistic, dynamical nature of our model means that there are
sufficient fluctuations present to enable oil and water particles from
the initially separated layers to move locally, and in so doing come
under the influence of the colour field gradients produced by other
layers of the same type. Since $\alpha = \beta = 1.0$, there is an
inherent tendency for the oil and water to act as immiscible fluids and
phase separate ({\it cf.}, Fig.~\ref{fig:tif}), and that is exactly what
the simulation, Fig.~\ref{fig:lms1}, reflects.

\begin{figure}
\begin{center}
\leavevmode
\hbox{
\epsfxsize=6.4in
\epsffile{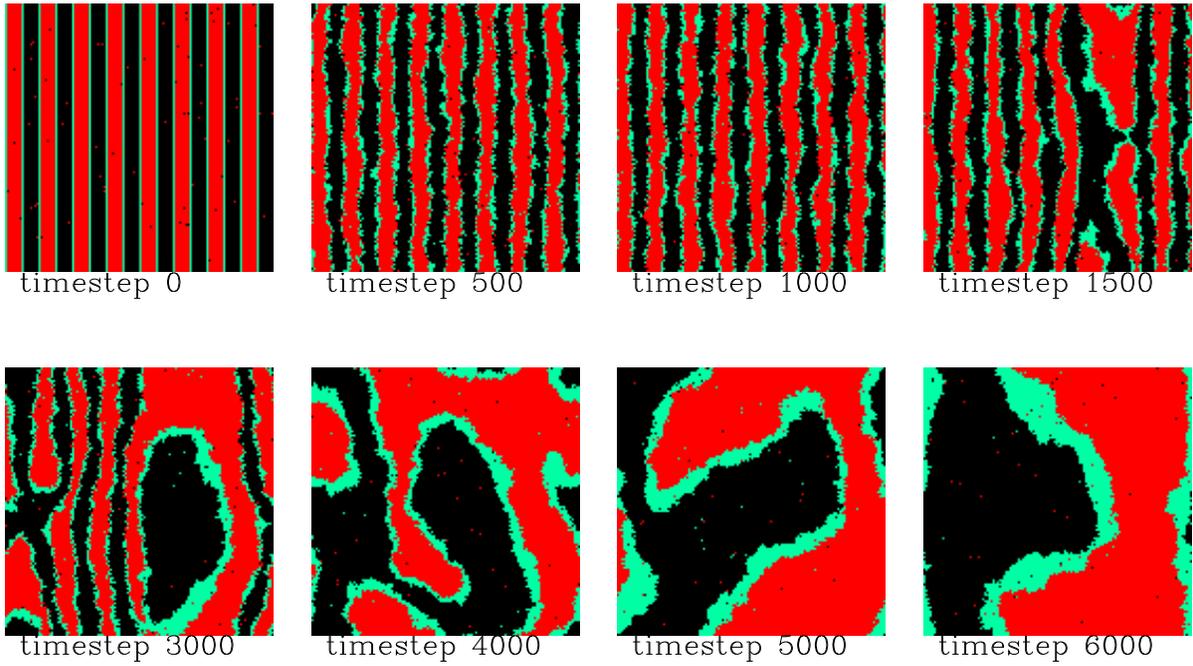}}
\end{center}
\caption{\sl Time evolution from initial lamellar configuration, for
the case with surfactant but only $\alpha$ non-zero.}
\label{fig:lms2}
\end{figure}

\begin{figure}
\begin{center}
\leavevmode
\hbox{
\epsfxsize=6.4in
\epsffile{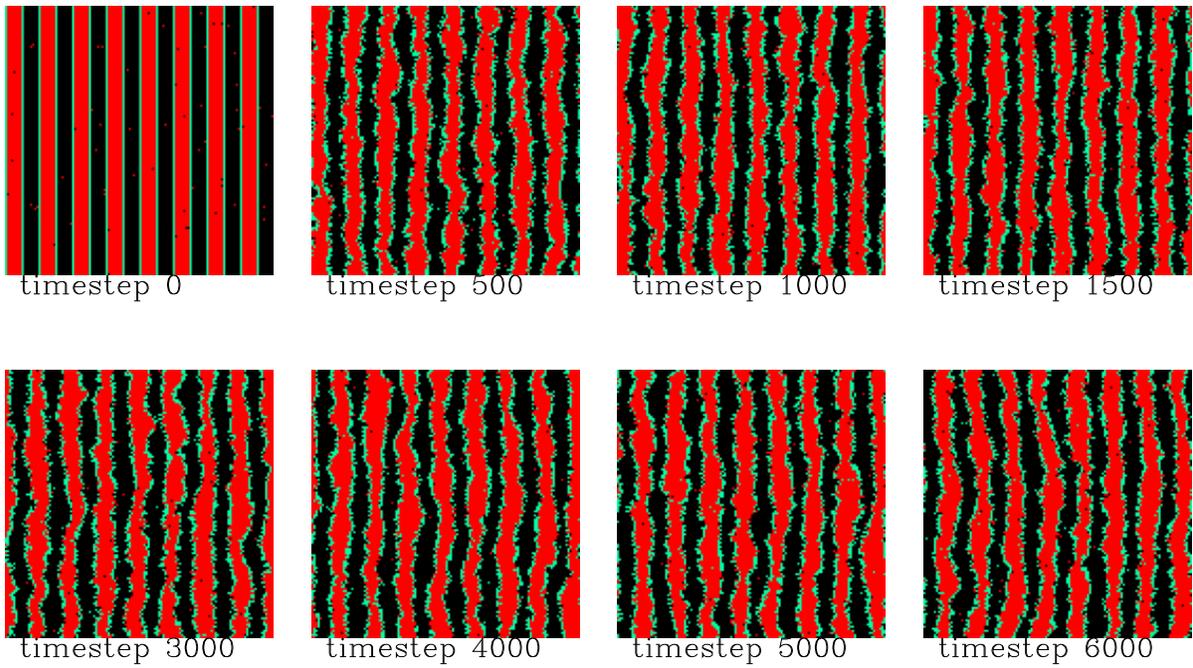}}
\end{center}
\caption{\sl Time evolution from initial lamellar configuration, for
the case with surfactant and coefficients set as $\alpha = 1.0, \mu =
0.05,  \epsilon = 8.0, \zeta = 0.5$.}
\label{fig:lms3}
\end{figure}
\newpage

The second result, displayed in Fig.~\ref{fig:lms2}, is included as a
control.  It shows that the mere presence of a third type of particle at
the oil-water interfaces is not enough to artificially stabilise the
structure of the layers. However it is clear that the initial breakup of
the layers takes place at a later time than for the case of two species
only (Fig.~\ref{fig:lms1}). This is because of the presence of a third
species between the oil and water layers, meaning that on average the
oil/water particles will have further to move before they can come under
the influence of a colour field set up by a neighbouring layer of like
colour. An interesting point to observe here is the forced accumulation
of the third species, in thick layers at the interfaces, once the oil
and water begin to act as immiscible fluids and phase separate.

Finally we come to Fig.~\ref{fig:lms3}. The presence of the colour
dipoles and the fact that their corresponding interaction terms can
affect the collision outcomes means, as the simulation clearly shows,
that it is now energetically favourable for the surfactant to remain at
oil-water interfaces. In spite of the fact that the dipolar vectors that
represent the surfactant molecules are initially assigned random angles
between $0$ and $2\pi$, so that again we have initial fluctuations in
the system, it is clear that the integrity of the lamellar structure is
maintained and appears to be completely stable, even at late times in
the simulation. At the oil-water interfaces the amphiphile sits in {\it
thin} layers ({\it cf.} Fig.~\ref{fig:lms2}), which compares well with
the knowledge that in real microemulsions amphiphile tends to reside in
monolayers between regions of oil and water. The greater the surface
tension that exists at the interfaces within a system, the more that
system will act to try and reduce the amount of interfacial area
present. Since in our model the presence of surfactant at the oil/water
interfaces results in stabilisation of the lamellar structure, and such
a structure has a large interfacial area, it appears that the surface
tension is indeed being reduced. This analysis and result provide
initial confirmation of microemulsion-like behaviour in our model.

\subsection{Ternary phases: Oil-in-water (water-in-oil) and bicontinuous
microemulsions}

Finally we use our model to simulate the different ternary {\it
microemulsion phases} that are possible in $2D$, namely, the
oil-in-water droplet and the bicontinuous phases.  (Note that since, at
present, our model treats oil and water molecules in symmetric fashion,
the oil-in-water phase could equally well represent a water-in-oil
phase.)  In reality the two distinct microemulsion phases will form if
there are the correct relative amounts of oil, water and surfactant in
the system at a given temperature. The oil-in-water droplet phase
typically consists of finely divided, spherical regions of oil, with
stabilising monolayers of surfactant surrounding them, sitting in the
bulk water background.  If one increases the relative amount of oil in
the system and there is sufficient amphiphile present, one will observe
the formation of mutually percolating tubular regions of oil and water,
with layers of surfactant sitting at the interfaces. In both these
cases, the equilibrium state does not correspond to complete separation
of the immiscible oil and water regions, but rather to complex
structures with very different characteristic length scales that form as
a result of the presence of amphiphile.

\begin{figure}
\begin{center}
\leavevmode
\hbox{
\epsfxsize=6.4in
\epsffile{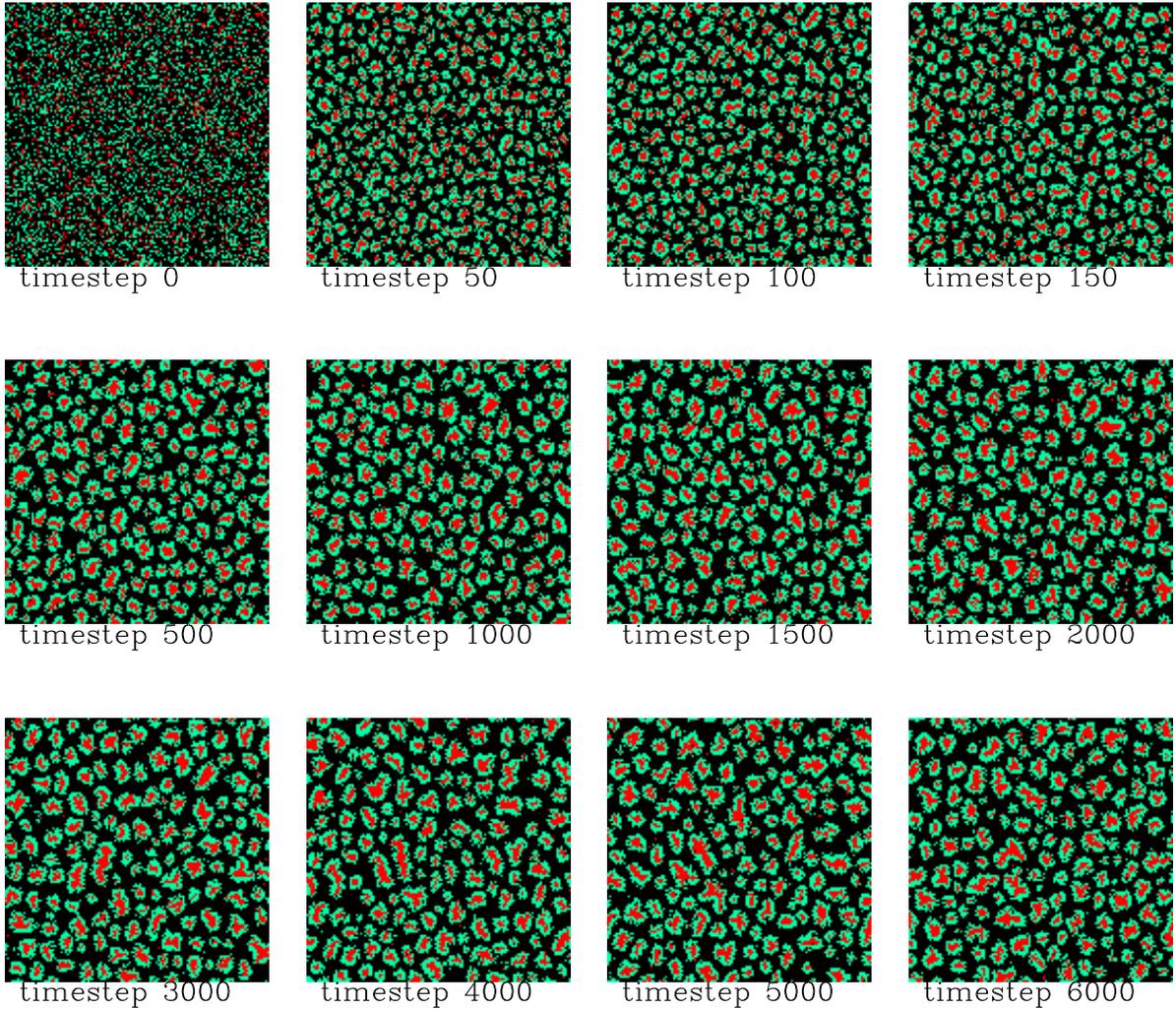}}
\end{center}
\caption{\sl Time evolution of oil-in-water microemulsion phase.}
\label{fig:me1}
\end{figure}

In order to reproduce the oil-in-water phase, we set up a simulation
with a random initial configuration consisting of a $3:1.9:0.7$
water-to-surfactant-to-oil ratio, respectively, with an averaged reduced
density of $0.56$. To maintain consistency between our various
simulations, we again use the coupling constants as defined in
Eq.~(\ref{eq:doc}).  Figure~\ref{fig:me1} displays the result. We see
the rapid formation of many oil-in-water droplets, whose size increases
slightly, but not without limit. This is representative of the
experimental microemulsion state and occurs because the layer of
surfactant that surrounds the droplets acts to stabilise the interfaces
and thereby inhibits the further flow of oil to the centre of any such
droplet, as well as discouraging their coalescence.

\begin{figure}
\begin{center}
\leavevmode
\hbox{
\epsfxsize=6.4in
\epsffile{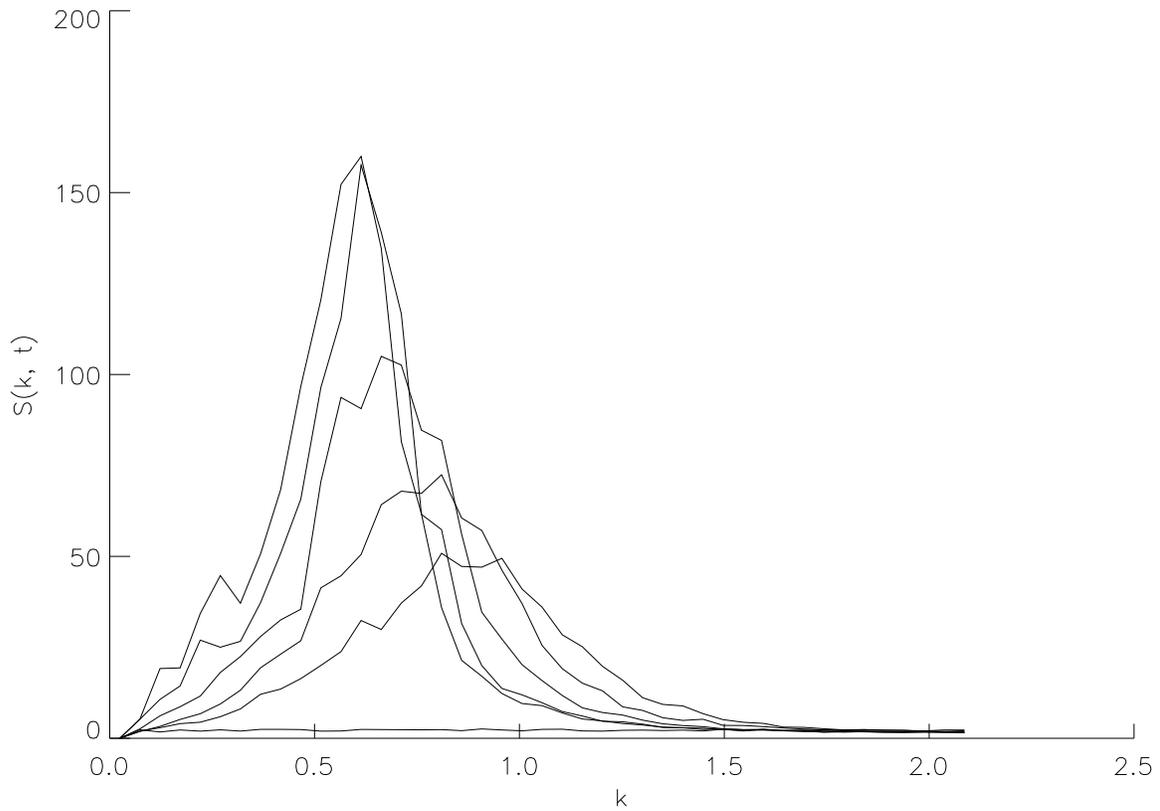}}
\end{center}
\caption{\sl Temporal evolution of $S(k, t)$ for microemulsion droplet case.
Timesteps shown are, from bottom to top, $t = 0, 40, 80, 200, 800,
2000$.}
\label{fig:medropsf}
\end{figure}

In order to quantify this result, we calculate the circularly averaged
structure factor of the colour charge, and plot the result in
Fig.~\ref{fig:medropsf}.  We observe the early-time growth of the peak
height of $S(k, t)$ as the peak itself shifts towards lower values of
the wavelength, indicating that the droplets form and grow to some
characteristic size. From at least timestep $800$ onwards there appears
to be a negligible amount of further growth or movement of the position
of the peak, indicating that the droplets have reached their maximum
size and will grow no more. This simple analysis confirms the ability of
our model to attain a microemulsion droplet phase within a certain
region of the overall phase diagram.

\begin{figure}
\begin{center}
\leavevmode
\hbox{
\epsfxsize=6.4in
\epsffile{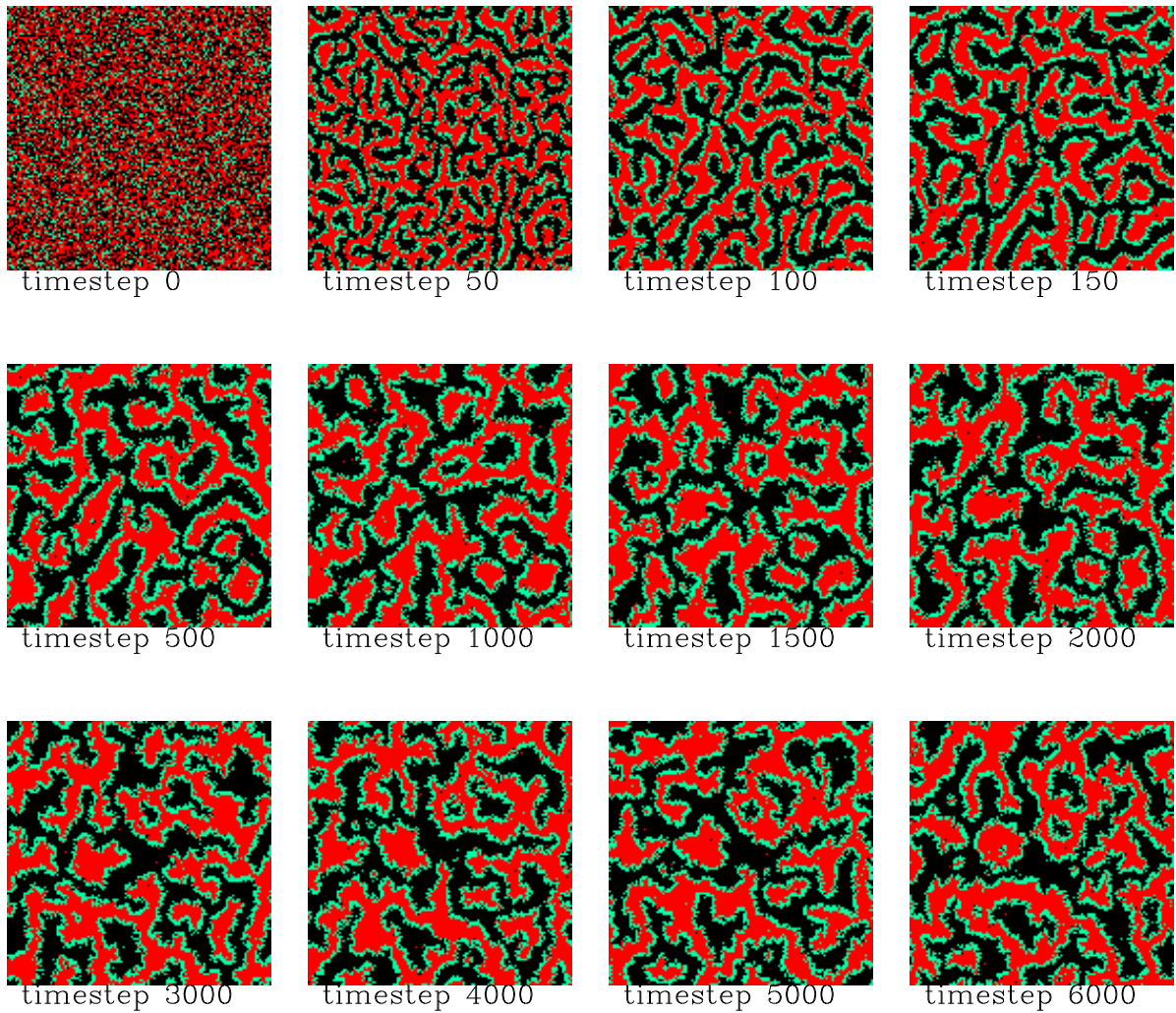}}
\end{center}
\caption{\sl Time evolution of bicontinuous microemulsion phase.}
\label{fig:me2}
\end{figure}

To demonstrate the existence of the bicontinuous regime within the
model's phase diagram, we use the same coupling constants as before but
simply increase the relative amount of oil present in the system. Hence
this second simulation, shown in Fig.~\ref{fig:me2}, has a random
initial mixture with a reduced density $0.55$ and a $3:2.25:3$
oil-to-surfactant-to-water ratio.

We observe the growth of an interconnected network of tubular-like
regions of oil and water, separated by {\it thin} layers of amphiphile
that ensure the stabilisation (reduction of oil-water surface tension)
of the bicontinuous r\'{e}gime, together with the formation of droplets
and some micelle-like objects. Note that on average the width of the oil
and water regions grows in size up to about $500$ timesteps, during
which time the surfactant particles migrate around the various oil-water
interfaces so as to spread themselves uniformly. Beyond this stage the
system changes very little, indicating that the observed bicontinuous
phase, although always slightly affected by the underlying lattice gas
dynamics, is stable.  To appreciate the significance of this result, the
snapshots and timescale should be compared with the two-immiscible-fluid
case (Fig.~\ref{fig:tif}), the only difference between the two being the
introduction of amphiphile and the accompanying interaction terms.

\begin{figure}
\begin{center}
\leavevmode
\hbox{
\epsfxsize=6.4in
\epsffile{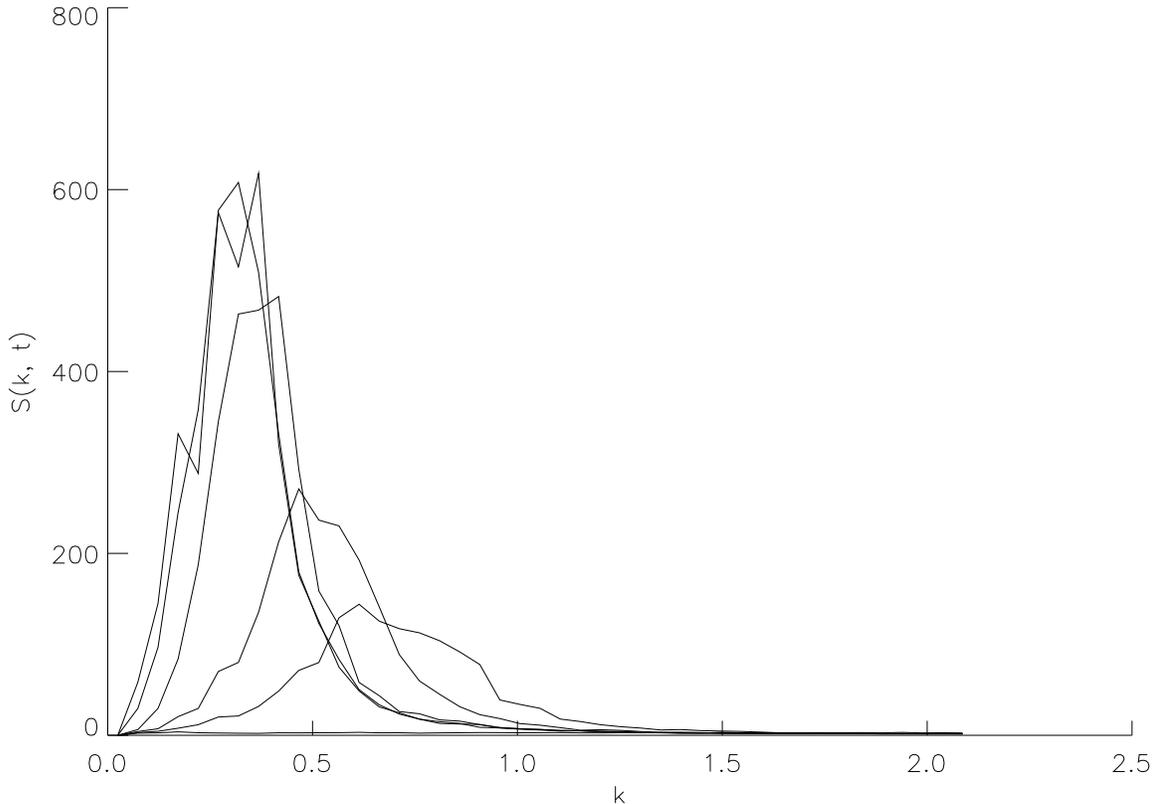}}
\end{center}
\caption{\sl Temporal evolution of $S(k, t)$ for bicontinuous
microemulsion case. Timesteps shown are, from bottom
to top, t = 0, 40, 80, 200, 800, 2000.}
\label{fig:mebicsf}
\end{figure}

To permit further analysis of this result, we calculate the circularly
averaged structure factor of the colour charge, in an exactly analagous
way to that for the immiscible fluid case.  The result we obtain is
shown in Fig.~\ref{fig:mebicsf}, which is an average over five
independent simulations.

Comparing this with Fig.~\ref{fig:sftif}, which shows the structure
factor for the immiscible oil and water case, we see that after about
$200$ timesteps the growth of the peak of the structure factor is
dramatically slowed down by the presence of the surfactant, and also
that some minimum $k$ is reached below which the location of the peak no
longer shifts, indicating that the underlying immiscible fluid
coalescence process has been inhibited owing to the presence of the
amphiphile.

\section{Discussion}

The results described above provide evidence of the ability of our model
to reproduce the fundamental equilibrium microemulsion phases.  Since
the dynamics conserves momentum in addition to mass, we expect that it
ought to be able to model dynamic, nonequilibrium behavior as well.  In
drawing this conclusion, however, an important caveat should be
mentioned: as noted earlier, lattice gases can break galilean
invariance, due to the presence of a preferred galilean frame of
reference, namely the lattice itself.  Mathematically, this problem
manifests itself by a spurious factor multiplying the inertial term in
the momentum-conserving Navier-Stokes equation.  For a single-phase
lattice gas, this factor can easily be scaled away; for compressible
flow, or for multiphase flow with interfaces, however, the presence of
this factor is problematic, and various techniques have been proposed to
remove it.  It has been shown that this can be done at the expense of
complicating the collision rules by introducing judicious violations of
semi-detailed balance (d'Humi\`{e}res {\it et. al.} 1987), by adding
many rest particles at each site (Gunstensen \& Rothman 1991b), or by
using multiple bits in each direction rather than only one (Boghosian
{\it et. al.} 1995).  In the present paper, we have evaded this issue by
focusing only on equilibrium phenomenology, or on creep flow situations
for which the inertial term is negligible.  For flow at finite Reynolds
number, however, this issue must be addressed using one of the
above-mentioned methods, and we plan to do this in future work.

In spite of this limitation, we note that the model exhibits some
interesting effects.  One of these, the roughening of the interface, is
evident in Fig.~\ref{fig:me2}.  The surfactant, in its attempt to
increase the surface area surrounding a given volume of oil, has caused
the creation of a fractal, fluctuating interface.  Lattice gases have
already been used as a tool for studying interface fluctuations in
immiscible fluids (Adler {\it et. al.} 1994), we now have the ability to
extend these studies to include the effect of surfactant on the
interface.

\section{Conclusions}
\label{sec:conc}

We have developed a model for momentum-conserving simulations of the
dynamics of microemulsions and other related self-assembling amphiphilic
systems. Using an electrostatic analogy for both the colour particles
and the amphiphilic colour dipoles we have been able to derive the
various energy interaction terms, including that of the Rothman-Keller
immiscible fluid lattice gas, from a microscopic particulate viewpoint.
Using a single set of coupling constants, we have shown that our model
exhibits the correct $2D$ phenomenology for both binary and ternary
phase systems using a combination of visual and analytic techniques;
various experimentally observed self-assembling structures form in a
consistent manner as a result of adjusting the relative amounts of oil,
water and amphiphile in the system.  The presence of enough surfactant
in the system clearly halts the phase separation of oil and water, and
this is acheived without altering the coupling constant $\alpha$ from a
value that produces immisible behaviour in the case of no surfactant.
In achieving these results, we have also demonstrated for the first time
that lattice gases may be used to investigate the dynamics of fluids
with very complex interactions.

Consequently we should be able to investigate a plethora of
microemulsion-related problems, including, for example, roughening and
interface fluctuations in microemulsions, and the behaviour of
microemulsions under flow conditions.  Such studies are fairly easily
implemented using lattice gases, and should also permit us to observe
the various microemulsion phases flowing through complex geometries such
as porous media.  Future extensions include a $3D$ version of our model,
which would allow various applications in many different areas of
science to be investigated.

\section*{Acknowledgments}
\label{sec:ack}

We are grateful to William Osborn, Dan Rothman, David Sherrington, Mike
Swift and Julia Yeomans for numerous helpful discussions during the
development of this work. We are indebted to Schlumberger Cambridge
Research and NATO for financial support for this project. One of us
(BMB) was supported in part by Phillips Laboratories and by the United
States Air Force Office of Scientific Research under grant number
F49620-95-1-0285; another of us (ANE) wishes to thank EPSRC and
Schlumberger for funding his CASE award.

\appendix
\section{Derivation of Equivalence Classes}
\label{sec:aeq}

As discussed in the main text, the partitioning of the $(M+1)^n$
possible states, in which each lattice site can exist, into equivalence
classes of states that have the same values for the $(M+D)$ conserved
quantities is required for the specification of the collision process.
Here we show how to derive those equivalence classes for lattice
collisions conserving both mass and momentum.  We deal specifically with
the case of three species ($M=3$) on a two-dimensional lattice ($D=2$)
so that we conserve three masses and two components of momentum.

Since we have seven lattice vectors per site (six plus the rest
particle) and each of these directions is assigned two bits, there is a
total of $n=14$ bits per lattice site and hence $2^{14}$ possible states
per site.  Note that the two bits correspond to no molecule ($00$), a
water molecule ($01$), a surfactant molecule ($10$), and an oil molecule
($11$).  We also need a way to identify the classes. We tag them by the
number of oil, water and surfactant particles, the x-momentum and the
y-momentum. Since there is a maximum of seven particles of any type at a
lattice site the masses can be stored using three bits for each, while
the $x$ and $y$ components of the momentum will need four bits and three
bits respectively as discussed below, giving a total of sixteen bits of
class information.

To see why the momentum components require four and three bits,
respectively, we orientate the hexagonal lattice so that the lattice
vectors are
\[
\hat{{\bf c}}_j =
 \left\{
  \begin{array}{ll}
  \hat{{\bf x}}\cos\left(\frac{2\pi j}{6}\right) +
  \hat{{\bf y}}\sin\left(\frac{2\pi j}{6}\right) &
  \mbox{for $0\leq j\leq 5$} \\
  {\bf 0} &
  \mbox{for $j=6.$}
  \end{array}
 \right.
\]
Defining $n_i\equiv\sum_\alpha n^\alpha_i$, the $x$-momentum is given
by,
\[
 p_x = n_0 + \frac{1}{2} n_1 -
       \frac{1}{2} n_2 - n_3 -
       \frac{1}{2} n_4 + \frac{1}{2} n_5.
\]
Since this can be a non-integer we multiply it by two to get a unique
integer identifier for $p_x$,
\bge
 2 p_x = 2n_0 + n_1 - n_2 - 2n_3 - n_4 + n_5. \label{eq:p_x}
\ee
We observe that this can be negative and that we would like an unsigned
integer to identify the x-momentum, so we make use of the fact that
$N$,
\[
 N = n_0 + n_1 +  n_2 + n_3 + n_4 + n_5 + n_6,
\]
is a conserved quantity, since it is just the total number of particles
at a site.  We then add twice $N$ to Eq.~(\ref{eq:p_x}) for $p_x$
to get,
\[
 2\left(N + p_x\right) = 4n_0 + 3n_1 + n_2 + n_4 + 3n_5 + 2n_6,
\]
which we can now use as an identifier for the x-momentum. Note that its
maximum value is $14$, so we need four bits to store this quantity.

Similarly the y-momentum is given by
\[
 p_y = \frac{\sqrt{3}}{2}(n_1 + n_2 - n_4 - n_5).
\]
In a similar way to that just described we multiply this equation by
$2/\sqrt{3}$ and
add $N$, getting a unique y-momentum identifier,
\[
\left( N + \frac{2}{\sqrt{3}}p_y\right) =
 n_0 + 2n_1 + 2n_2 + n_3 + n_6.
\]
This has a maximum value of $7$, so it can be stored in only
$3$ bits.

We can now loop through all states and obtain the $16$-bit class
identifier for each one. These class identifiers are then sorted so that
states in the same equivalence class are next to each other. The states
and class identifiers can subsequently be separated into two
corresponding arrays. We use the $14$-bit state as a key to look up the
initial index and length of that state's equivalence class. All three
look-up tables, the class-pointers, class-lengths and states are then
integer arrays of length $2^{14}$ which can be precomputed and easily
read by the main program code at run time.

\section{Sampling Procedure}
\label{sec:asp}

We can sample the outgoing occupation numbers $\bfn'$ directly from
Eq.~(\ref{eq:nn6}).  This is accomplished by obtaining the total energy
value for each outgoing state in the particular equivalence class,
normalising these energies and then making use of a histogram
probability distribution method to select the outgoing state.

\begin{figure}
\setlength{\unitlength}{0.240900pt}
\ifx\plotpoint\undefined\newsavebox{\plotpoint}\fi
\sbox{\plotpoint}{\rule[-0.175pt]{0.350pt}{0.350pt}}%
\begin{picture}(1500,900)(0,0)
\tenrm
\put(264,158){\rule[-0.175pt]{282.335pt}{0.350pt}}
\put(264,158){\rule[-0.175pt]{0.350pt}{151.526pt}}
\put(264,158){\rule[-0.175pt]{4.818pt}{0.350pt}}
\put(264,327){\rule[-0.175pt]{4.818pt}{0.350pt}}
\put(242,327){\makebox(0,0)[r]{$\frac{\exp(-z)}{2\pi I_0(z)}$}}
\put(1416,327){\rule[-0.175pt]{4.818pt}{0.350pt}}
\put(264,618){\rule[-0.175pt]{4.818pt}{0.350pt}}
\put(242,618){\makebox(0,0)[r]{$\frac{\exp(z)}{2\pi I_0(z)}$}}
\put(1416,618){\rule[-0.175pt]{4.818pt}{0.350pt}}
\put(242,810){\makebox(0,0)[r]{$Q(\theta^{\prime\prime})$}}
\put(850,20){\makebox(0,0){$\theta^{\prime\prime}$}}
\put(264,158){\rule[-0.175pt]{0.350pt}{4.818pt}}
\put(264,113){\makebox(0,0){0}}
\put(264,767){\rule[-0.175pt]{0.350pt}{4.818pt}}
\put(850,158){\rule[-0.175pt]{0.350pt}{4.818pt}}
\put(850,113){\makebox(0,0){$\pi$}}
\put(850,767){\rule[-0.175pt]{0.350pt}{4.818pt}}
\put(1436,158){\rule[-0.175pt]{0.350pt}{4.818pt}}
\put(1436,113){\makebox(0,0){$2\pi$}}
\put(1436,767){\rule[-0.175pt]{0.350pt}{4.818pt}}
\put(264,158){\rule[-0.175pt]{282.335pt}{0.350pt}}
\put(1436,158){\rule[-0.175pt]{0.350pt}{151.526pt}}
\put(264,787){\rule[-0.175pt]{282.335pt}{0.350pt}}
\put(264,158){\rule[-0.175pt]{0.350pt}{151.526pt}}
\put(264,327){\usebox{\plotpoint}}
\put(264,327){\rule[-0.175pt]{5.782pt}{0.350pt}}
\put(288,328){\rule[-0.175pt]{2.891pt}{0.350pt}}
\put(300,329){\rule[-0.175pt]{2.650pt}{0.350pt}}
\put(311,330){\rule[-0.175pt]{2.891pt}{0.350pt}}
\put(323,331){\rule[-0.175pt]{1.445pt}{0.350pt}}
\put(329,332){\rule[-0.175pt]{1.445pt}{0.350pt}}
\put(335,333){\rule[-0.175pt]{0.964pt}{0.350pt}}
\put(339,334){\rule[-0.175pt]{0.964pt}{0.350pt}}
\put(343,335){\rule[-0.175pt]{0.964pt}{0.350pt}}
\put(347,336){\rule[-0.175pt]{1.445pt}{0.350pt}}
\put(353,337){\rule[-0.175pt]{1.445pt}{0.350pt}}
\put(359,338){\rule[-0.175pt]{0.964pt}{0.350pt}}
\put(363,339){\rule[-0.175pt]{0.964pt}{0.350pt}}
\put(367,340){\rule[-0.175pt]{0.964pt}{0.350pt}}
\put(371,341){\rule[-0.175pt]{0.883pt}{0.350pt}}
\put(374,342){\rule[-0.175pt]{0.883pt}{0.350pt}}
\put(378,343){\rule[-0.175pt]{0.883pt}{0.350pt}}
\put(381,344){\rule[-0.175pt]{0.723pt}{0.350pt}}
\put(385,345){\rule[-0.175pt]{0.723pt}{0.350pt}}
\put(388,346){\rule[-0.175pt]{0.723pt}{0.350pt}}
\put(391,347){\rule[-0.175pt]{0.723pt}{0.350pt}}
\put(394,348){\rule[-0.175pt]{0.723pt}{0.350pt}}
\put(397,349){\rule[-0.175pt]{0.723pt}{0.350pt}}
\put(400,350){\rule[-0.175pt]{0.723pt}{0.350pt}}
\put(403,351){\rule[-0.175pt]{0.723pt}{0.350pt}}
\put(406,352){\rule[-0.175pt]{0.578pt}{0.350pt}}
\put(408,353){\rule[-0.175pt]{0.578pt}{0.350pt}}
\put(410,354){\rule[-0.175pt]{0.578pt}{0.350pt}}
\put(413,355){\rule[-0.175pt]{0.578pt}{0.350pt}}
\put(415,356){\rule[-0.175pt]{0.578pt}{0.350pt}}
\put(417,357){\rule[-0.175pt]{0.723pt}{0.350pt}}
\put(421,358){\rule[-0.175pt]{0.723pt}{0.350pt}}
\put(424,359){\rule[-0.175pt]{0.723pt}{0.350pt}}
\put(427,360){\rule[-0.175pt]{0.723pt}{0.350pt}}
\put(430,361){\rule[-0.175pt]{0.482pt}{0.350pt}}
\put(432,362){\rule[-0.175pt]{0.482pt}{0.350pt}}
\put(434,363){\rule[-0.175pt]{0.482pt}{0.350pt}}
\put(436,364){\rule[-0.175pt]{0.482pt}{0.350pt}}
\put(438,365){\rule[-0.175pt]{0.482pt}{0.350pt}}
\put(440,366){\rule[-0.175pt]{0.482pt}{0.350pt}}
\put(442,367){\rule[-0.175pt]{0.530pt}{0.350pt}}
\put(444,368){\rule[-0.175pt]{0.530pt}{0.350pt}}
\put(446,369){\rule[-0.175pt]{0.530pt}{0.350pt}}
\put(448,370){\rule[-0.175pt]{0.530pt}{0.350pt}}
\put(450,371){\rule[-0.175pt]{0.530pt}{0.350pt}}
\put(453,372){\rule[-0.175pt]{0.482pt}{0.350pt}}
\put(455,373){\rule[-0.175pt]{0.482pt}{0.350pt}}
\put(457,374){\rule[-0.175pt]{0.482pt}{0.350pt}}
\put(459,375){\rule[-0.175pt]{0.482pt}{0.350pt}}
\put(461,376){\rule[-0.175pt]{0.482pt}{0.350pt}}
\put(463,377){\rule[-0.175pt]{0.482pt}{0.350pt}}
\put(465,378){\rule[-0.175pt]{0.413pt}{0.350pt}}
\put(466,379){\rule[-0.175pt]{0.413pt}{0.350pt}}
\put(468,380){\rule[-0.175pt]{0.413pt}{0.350pt}}
\put(470,381){\rule[-0.175pt]{0.413pt}{0.350pt}}
\put(471,382){\rule[-0.175pt]{0.413pt}{0.350pt}}
\put(473,383){\rule[-0.175pt]{0.413pt}{0.350pt}}
\put(475,384){\rule[-0.175pt]{0.413pt}{0.350pt}}
\put(477,385){\rule[-0.175pt]{0.482pt}{0.350pt}}
\put(479,386){\rule[-0.175pt]{0.482pt}{0.350pt}}
\put(481,387){\rule[-0.175pt]{0.482pt}{0.350pt}}
\put(483,388){\rule[-0.175pt]{0.482pt}{0.350pt}}
\put(485,389){\rule[-0.175pt]{0.482pt}{0.350pt}}
\put(487,390){\rule[-0.175pt]{0.482pt}{0.350pt}}
\put(489,391){\rule[-0.175pt]{0.413pt}{0.350pt}}
\put(490,392){\rule[-0.175pt]{0.413pt}{0.350pt}}
\put(492,393){\rule[-0.175pt]{0.413pt}{0.350pt}}
\put(494,394){\rule[-0.175pt]{0.413pt}{0.350pt}}
\put(495,395){\rule[-0.175pt]{0.413pt}{0.350pt}}
\put(497,396){\rule[-0.175pt]{0.413pt}{0.350pt}}
\put(499,397){\rule[-0.175pt]{0.413pt}{0.350pt}}
\put(501,398){\rule[-0.175pt]{0.361pt}{0.350pt}}
\put(502,399){\rule[-0.175pt]{0.361pt}{0.350pt}}
\put(504,400){\rule[-0.175pt]{0.361pt}{0.350pt}}
\put(505,401){\rule[-0.175pt]{0.361pt}{0.350pt}}
\put(507,402){\rule[-0.175pt]{0.361pt}{0.350pt}}
\put(508,403){\rule[-0.175pt]{0.361pt}{0.350pt}}
\put(510,404){\rule[-0.175pt]{0.361pt}{0.350pt}}
\put(511,405){\rule[-0.175pt]{0.361pt}{0.350pt}}
\put(513,406){\usebox{\plotpoint}}
\put(514,407){\usebox{\plotpoint}}
\put(515,408){\usebox{\plotpoint}}
\put(517,409){\usebox{\plotpoint}}
\put(518,410){\usebox{\plotpoint}}
\put(519,411){\usebox{\plotpoint}}
\put(521,412){\usebox{\plotpoint}}
\put(522,413){\usebox{\plotpoint}}
\put(524,414){\rule[-0.175pt]{0.361pt}{0.350pt}}
\put(525,415){\rule[-0.175pt]{0.361pt}{0.350pt}}
\put(527,416){\rule[-0.175pt]{0.361pt}{0.350pt}}
\put(528,417){\rule[-0.175pt]{0.361pt}{0.350pt}}
\put(530,418){\rule[-0.175pt]{0.361pt}{0.350pt}}
\put(531,419){\rule[-0.175pt]{0.361pt}{0.350pt}}
\put(533,420){\rule[-0.175pt]{0.361pt}{0.350pt}}
\put(534,421){\rule[-0.175pt]{0.361pt}{0.350pt}}
\put(536,422){\rule[-0.175pt]{0.361pt}{0.350pt}}
\put(537,423){\rule[-0.175pt]{0.361pt}{0.350pt}}
\put(539,424){\rule[-0.175pt]{0.361pt}{0.350pt}}
\put(540,425){\rule[-0.175pt]{0.361pt}{0.350pt}}
\put(542,426){\rule[-0.175pt]{0.361pt}{0.350pt}}
\put(543,427){\rule[-0.175pt]{0.361pt}{0.350pt}}
\put(545,428){\rule[-0.175pt]{0.361pt}{0.350pt}}
\put(546,429){\rule[-0.175pt]{0.361pt}{0.350pt}}
\put(548,430){\usebox{\plotpoint}}
\put(549,431){\usebox{\plotpoint}}
\put(550,432){\usebox{\plotpoint}}
\put(551,433){\usebox{\plotpoint}}
\put(553,434){\usebox{\plotpoint}}
\put(554,435){\usebox{\plotpoint}}
\put(555,436){\usebox{\plotpoint}}
\put(557,437){\usebox{\plotpoint}}
\put(558,438){\usebox{\plotpoint}}
\put(559,439){\usebox{\plotpoint}}
\put(561,440){\usebox{\plotpoint}}
\put(562,441){\usebox{\plotpoint}}
\put(563,442){\usebox{\plotpoint}}
\put(565,443){\usebox{\plotpoint}}
\put(566,444){\usebox{\plotpoint}}
\put(567,445){\usebox{\plotpoint}}
\put(569,446){\usebox{\plotpoint}}
\put(570,447){\usebox{\plotpoint}}
\put(571,448){\usebox{\plotpoint}}
\put(573,449){\usebox{\plotpoint}}
\put(574,450){\usebox{\plotpoint}}
\put(575,451){\usebox{\plotpoint}}
\put(577,452){\usebox{\plotpoint}}
\put(578,453){\usebox{\plotpoint}}
\put(579,454){\usebox{\plotpoint}}
\put(581,455){\usebox{\plotpoint}}
\put(582,456){\usebox{\plotpoint}}
\put(583,457){\usebox{\plotpoint}}
\put(585,458){\usebox{\plotpoint}}
\put(586,459){\usebox{\plotpoint}}
\put(587,460){\usebox{\plotpoint}}
\put(588,461){\usebox{\plotpoint}}
\put(589,462){\usebox{\plotpoint}}
\put(590,463){\usebox{\plotpoint}}
\put(591,464){\usebox{\plotpoint}}
\put(592,465){\usebox{\plotpoint}}
\put(593,466){\usebox{\plotpoint}}
\put(594,467){\usebox{\plotpoint}}
\put(596,468){\usebox{\plotpoint}}
\put(597,469){\usebox{\plotpoint}}
\put(598,470){\usebox{\plotpoint}}
\put(599,471){\usebox{\plotpoint}}
\put(601,472){\usebox{\plotpoint}}
\put(602,473){\usebox{\plotpoint}}
\put(603,474){\usebox{\plotpoint}}
\put(604,475){\usebox{\plotpoint}}
\put(605,476){\usebox{\plotpoint}}
\put(607,477){\usebox{\plotpoint}}
\put(608,478){\usebox{\plotpoint}}
\put(609,479){\usebox{\plotpoint}}
\put(610,480){\usebox{\plotpoint}}
\put(612,481){\usebox{\plotpoint}}
\put(613,482){\usebox{\plotpoint}}
\put(614,483){\usebox{\plotpoint}}
\put(616,484){\usebox{\plotpoint}}
\put(617,485){\usebox{\plotpoint}}
\put(618,486){\usebox{\plotpoint}}
\put(620,487){\usebox{\plotpoint}}
\put(621,488){\usebox{\plotpoint}}
\put(622,489){\usebox{\plotpoint}}
\put(623,490){\usebox{\plotpoint}}
\put(625,491){\usebox{\plotpoint}}
\put(626,492){\usebox{\plotpoint}}
\put(627,493){\usebox{\plotpoint}}
\put(628,494){\usebox{\plotpoint}}
\put(629,495){\usebox{\plotpoint}}
\put(631,496){\usebox{\plotpoint}}
\put(632,497){\usebox{\plotpoint}}
\put(633,498){\usebox{\plotpoint}}
\put(634,499){\usebox{\plotpoint}}
\put(635,500){\usebox{\plotpoint}}
\put(637,501){\usebox{\plotpoint}}
\put(638,502){\usebox{\plotpoint}}
\put(639,503){\usebox{\plotpoint}}
\put(640,504){\usebox{\plotpoint}}
\put(641,505){\usebox{\plotpoint}}
\put(643,506){\usebox{\plotpoint}}
\put(644,507){\usebox{\plotpoint}}
\put(645,508){\usebox{\plotpoint}}
\put(646,509){\usebox{\plotpoint}}
\put(647,510){\usebox{\plotpoint}}
\put(649,511){\usebox{\plotpoint}}
\put(650,512){\usebox{\plotpoint}}
\put(651,513){\usebox{\plotpoint}}
\put(652,514){\usebox{\plotpoint}}
\put(653,515){\usebox{\plotpoint}}
\put(655,516){\usebox{\plotpoint}}
\put(656,517){\usebox{\plotpoint}}
\put(657,518){\usebox{\plotpoint}}
\put(658,519){\usebox{\plotpoint}}
\put(659,520){\usebox{\plotpoint}}
\put(661,521){\usebox{\plotpoint}}
\put(662,522){\usebox{\plotpoint}}
\put(663,523){\usebox{\plotpoint}}
\put(664,524){\usebox{\plotpoint}}
\put(665,525){\usebox{\plotpoint}}
\put(667,526){\usebox{\plotpoint}}
\put(668,527){\usebox{\plotpoint}}
\put(669,528){\usebox{\plotpoint}}
\put(670,529){\usebox{\plotpoint}}
\put(671,530){\usebox{\plotpoint}}
\put(672,531){\usebox{\plotpoint}}
\put(673,532){\usebox{\plotpoint}}
\put(674,533){\usebox{\plotpoint}}
\put(675,534){\usebox{\plotpoint}}
\put(676,535){\usebox{\plotpoint}}
\put(677,536){\usebox{\plotpoint}}
\put(679,537){\usebox{\plotpoint}}
\put(680,538){\usebox{\plotpoint}}
\put(681,539){\usebox{\plotpoint}}
\put(683,540){\usebox{\plotpoint}}
\put(684,541){\usebox{\plotpoint}}
\put(685,542){\usebox{\plotpoint}}
\put(687,543){\usebox{\plotpoint}}
\put(688,544){\usebox{\plotpoint}}
\put(689,545){\usebox{\plotpoint}}
\put(691,546){\usebox{\plotpoint}}
\put(692,547){\usebox{\plotpoint}}
\put(693,548){\usebox{\plotpoint}}
\put(695,549){\usebox{\plotpoint}}
\put(696,550){\usebox{\plotpoint}}
\put(697,551){\usebox{\plotpoint}}
\put(699,552){\usebox{\plotpoint}}
\put(700,553){\usebox{\plotpoint}}
\put(701,554){\usebox{\plotpoint}}
\put(703,555){\usebox{\plotpoint}}
\put(704,556){\usebox{\plotpoint}}
\put(705,557){\usebox{\plotpoint}}
\put(707,558){\usebox{\plotpoint}}
\put(708,559){\usebox{\plotpoint}}
\put(709,560){\usebox{\plotpoint}}
\put(711,561){\usebox{\plotpoint}}
\put(712,562){\usebox{\plotpoint}}
\put(713,563){\rule[-0.175pt]{0.361pt}{0.350pt}}
\put(715,564){\rule[-0.175pt]{0.361pt}{0.350pt}}
\put(717,565){\rule[-0.175pt]{0.361pt}{0.350pt}}
\put(718,566){\rule[-0.175pt]{0.361pt}{0.350pt}}
\put(720,567){\rule[-0.175pt]{0.361pt}{0.350pt}}
\put(721,568){\rule[-0.175pt]{0.361pt}{0.350pt}}
\put(723,569){\rule[-0.175pt]{0.361pt}{0.350pt}}
\put(724,570){\rule[-0.175pt]{0.361pt}{0.350pt}}
\put(726,571){\rule[-0.175pt]{0.361pt}{0.350pt}}
\put(727,572){\rule[-0.175pt]{0.361pt}{0.350pt}}
\put(729,573){\rule[-0.175pt]{0.361pt}{0.350pt}}
\put(730,574){\rule[-0.175pt]{0.361pt}{0.350pt}}
\put(732,575){\rule[-0.175pt]{0.361pt}{0.350pt}}
\put(733,576){\rule[-0.175pt]{0.361pt}{0.350pt}}
\put(735,577){\rule[-0.175pt]{0.361pt}{0.350pt}}
\put(736,578){\rule[-0.175pt]{0.361pt}{0.350pt}}
\put(738,579){\rule[-0.175pt]{0.379pt}{0.350pt}}
\put(739,580){\rule[-0.175pt]{0.379pt}{0.350pt}}
\put(741,581){\rule[-0.175pt]{0.379pt}{0.350pt}}
\put(742,582){\rule[-0.175pt]{0.379pt}{0.350pt}}
\put(744,583){\rule[-0.175pt]{0.379pt}{0.350pt}}
\put(745,584){\rule[-0.175pt]{0.379pt}{0.350pt}}
\put(747,585){\rule[-0.175pt]{0.379pt}{0.350pt}}
\put(748,586){\rule[-0.175pt]{0.413pt}{0.350pt}}
\put(750,587){\rule[-0.175pt]{0.413pt}{0.350pt}}
\put(752,588){\rule[-0.175pt]{0.413pt}{0.350pt}}
\put(754,589){\rule[-0.175pt]{0.413pt}{0.350pt}}
\put(755,590){\rule[-0.175pt]{0.413pt}{0.350pt}}
\put(757,591){\rule[-0.175pt]{0.413pt}{0.350pt}}
\put(759,592){\rule[-0.175pt]{0.413pt}{0.350pt}}
\put(761,593){\rule[-0.175pt]{0.482pt}{0.350pt}}
\put(763,594){\rule[-0.175pt]{0.482pt}{0.350pt}}
\put(765,595){\rule[-0.175pt]{0.482pt}{0.350pt}}
\put(767,596){\rule[-0.175pt]{0.482pt}{0.350pt}}
\put(769,597){\rule[-0.175pt]{0.482pt}{0.350pt}}
\put(771,598){\rule[-0.175pt]{0.482pt}{0.350pt}}
\put(773,599){\rule[-0.175pt]{0.578pt}{0.350pt}}
\put(775,600){\rule[-0.175pt]{0.578pt}{0.350pt}}
\put(777,601){\rule[-0.175pt]{0.578pt}{0.350pt}}
\put(780,602){\rule[-0.175pt]{0.578pt}{0.350pt}}
\put(782,603){\rule[-0.175pt]{0.578pt}{0.350pt}}
\put(785,604){\rule[-0.175pt]{0.578pt}{0.350pt}}
\put(787,605){\rule[-0.175pt]{0.578pt}{0.350pt}}
\put(789,606){\rule[-0.175pt]{0.578pt}{0.350pt}}
\put(792,607){\rule[-0.175pt]{0.578pt}{0.350pt}}
\put(794,608){\rule[-0.175pt]{0.578pt}{0.350pt}}
\put(797,609){\rule[-0.175pt]{0.964pt}{0.350pt}}
\put(801,610){\rule[-0.175pt]{0.964pt}{0.350pt}}
\put(805,611){\rule[-0.175pt]{0.964pt}{0.350pt}}
\put(809,612){\rule[-0.175pt]{0.883pt}{0.350pt}}
\put(812,613){\rule[-0.175pt]{0.883pt}{0.350pt}}
\put(816,614){\rule[-0.175pt]{0.883pt}{0.350pt}}
\put(820,615){\rule[-0.175pt]{1.445pt}{0.350pt}}
\put(826,616){\rule[-0.175pt]{1.445pt}{0.350pt}}
\put(832,617){\rule[-0.175pt]{2.891pt}{0.350pt}}
\put(844,618){\rule[-0.175pt]{5.782pt}{0.350pt}}
\put(868,617){\rule[-0.175pt]{1.445pt}{0.350pt}}
\put(874,616){\rule[-0.175pt]{1.445pt}{0.350pt}}
\put(880,615){\rule[-0.175pt]{0.883pt}{0.350pt}}
\put(883,614){\rule[-0.175pt]{0.883pt}{0.350pt}}
\put(887,613){\rule[-0.175pt]{0.883pt}{0.350pt}}
\put(891,612){\rule[-0.175pt]{0.964pt}{0.350pt}}
\put(895,611){\rule[-0.175pt]{0.964pt}{0.350pt}}
\put(899,610){\rule[-0.175pt]{0.964pt}{0.350pt}}
\put(903,609){\rule[-0.175pt]{0.578pt}{0.350pt}}
\put(905,608){\rule[-0.175pt]{0.578pt}{0.350pt}}
\put(907,607){\rule[-0.175pt]{0.578pt}{0.350pt}}
\put(910,606){\rule[-0.175pt]{0.578pt}{0.350pt}}
\put(912,605){\rule[-0.175pt]{0.578pt}{0.350pt}}
\put(915,604){\rule[-0.175pt]{0.578pt}{0.350pt}}
\put(917,603){\rule[-0.175pt]{0.578pt}{0.350pt}}
\put(919,602){\rule[-0.175pt]{0.578pt}{0.350pt}}
\put(922,601){\rule[-0.175pt]{0.578pt}{0.350pt}}
\put(924,600){\rule[-0.175pt]{0.578pt}{0.350pt}}
\put(927,599){\rule[-0.175pt]{0.482pt}{0.350pt}}
\put(929,598){\rule[-0.175pt]{0.482pt}{0.350pt}}
\put(931,597){\rule[-0.175pt]{0.482pt}{0.350pt}}
\put(933,596){\rule[-0.175pt]{0.482pt}{0.350pt}}
\put(935,595){\rule[-0.175pt]{0.482pt}{0.350pt}}
\put(937,594){\rule[-0.175pt]{0.482pt}{0.350pt}}
\put(939,593){\rule[-0.175pt]{0.413pt}{0.350pt}}
\put(940,592){\rule[-0.175pt]{0.413pt}{0.350pt}}
\put(942,591){\rule[-0.175pt]{0.413pt}{0.350pt}}
\put(944,590){\rule[-0.175pt]{0.413pt}{0.350pt}}
\put(945,589){\rule[-0.175pt]{0.413pt}{0.350pt}}
\put(947,588){\rule[-0.175pt]{0.413pt}{0.350pt}}
\put(949,587){\rule[-0.175pt]{0.413pt}{0.350pt}}
\put(951,586){\rule[-0.175pt]{0.379pt}{0.350pt}}
\put(952,585){\rule[-0.175pt]{0.379pt}{0.350pt}}
\put(954,584){\rule[-0.175pt]{0.379pt}{0.350pt}}
\put(955,583){\rule[-0.175pt]{0.379pt}{0.350pt}}
\put(957,582){\rule[-0.175pt]{0.379pt}{0.350pt}}
\put(958,581){\rule[-0.175pt]{0.379pt}{0.350pt}}
\put(960,580){\rule[-0.175pt]{0.379pt}{0.350pt}}
\put(961,579){\rule[-0.175pt]{0.361pt}{0.350pt}}
\put(963,578){\rule[-0.175pt]{0.361pt}{0.350pt}}
\put(965,577){\rule[-0.175pt]{0.361pt}{0.350pt}}
\put(966,576){\rule[-0.175pt]{0.361pt}{0.350pt}}
\put(968,575){\rule[-0.175pt]{0.361pt}{0.350pt}}
\put(969,574){\rule[-0.175pt]{0.361pt}{0.350pt}}
\put(971,573){\rule[-0.175pt]{0.361pt}{0.350pt}}
\put(972,572){\rule[-0.175pt]{0.361pt}{0.350pt}}
\put(974,571){\rule[-0.175pt]{0.361pt}{0.350pt}}
\put(975,570){\rule[-0.175pt]{0.361pt}{0.350pt}}
\put(977,569){\rule[-0.175pt]{0.361pt}{0.350pt}}
\put(978,568){\rule[-0.175pt]{0.361pt}{0.350pt}}
\put(980,567){\rule[-0.175pt]{0.361pt}{0.350pt}}
\put(981,566){\rule[-0.175pt]{0.361pt}{0.350pt}}
\put(983,565){\rule[-0.175pt]{0.361pt}{0.350pt}}
\put(984,564){\rule[-0.175pt]{0.361pt}{0.350pt}}
\put(986,563){\usebox{\plotpoint}}
\put(987,562){\usebox{\plotpoint}}
\put(988,561){\usebox{\plotpoint}}
\put(989,560){\usebox{\plotpoint}}
\put(991,559){\usebox{\plotpoint}}
\put(992,558){\usebox{\plotpoint}}
\put(993,557){\usebox{\plotpoint}}
\put(995,556){\usebox{\plotpoint}}
\put(996,555){\usebox{\plotpoint}}
\put(997,554){\usebox{\plotpoint}}
\put(999,553){\usebox{\plotpoint}}
\put(1000,552){\usebox{\plotpoint}}
\put(1001,551){\usebox{\plotpoint}}
\put(1003,550){\usebox{\plotpoint}}
\put(1004,549){\usebox{\plotpoint}}
\put(1005,548){\usebox{\plotpoint}}
\put(1007,547){\usebox{\plotpoint}}
\put(1008,546){\usebox{\plotpoint}}
\put(1009,545){\usebox{\plotpoint}}
\put(1011,544){\usebox{\plotpoint}}
\put(1012,543){\usebox{\plotpoint}}
\put(1013,542){\usebox{\plotpoint}}
\put(1015,541){\usebox{\plotpoint}}
\put(1016,540){\usebox{\plotpoint}}
\put(1017,539){\usebox{\plotpoint}}
\put(1019,538){\usebox{\plotpoint}}
\put(1020,537){\usebox{\plotpoint}}
\put(1021,536){\usebox{\plotpoint}}
\put(1023,535){\usebox{\plotpoint}}
\put(1024,534){\usebox{\plotpoint}}
\put(1025,533){\usebox{\plotpoint}}
\put(1026,532){\usebox{\plotpoint}}
\put(1027,531){\usebox{\plotpoint}}
\put(1028,530){\usebox{\plotpoint}}
\put(1029,529){\usebox{\plotpoint}}
\put(1030,528){\usebox{\plotpoint}}
\put(1031,527){\usebox{\plotpoint}}
\put(1032,526){\usebox{\plotpoint}}
\put(1034,525){\usebox{\plotpoint}}
\put(1035,524){\usebox{\plotpoint}}
\put(1036,523){\usebox{\plotpoint}}
\put(1037,522){\usebox{\plotpoint}}
\put(1038,521){\usebox{\plotpoint}}
\put(1040,520){\usebox{\plotpoint}}
\put(1041,519){\usebox{\plotpoint}}
\put(1042,518){\usebox{\plotpoint}}
\put(1043,517){\usebox{\plotpoint}}
\put(1044,516){\usebox{\plotpoint}}
\put(1046,515){\usebox{\plotpoint}}
\put(1047,514){\usebox{\plotpoint}}
\put(1048,513){\usebox{\plotpoint}}
\put(1049,512){\usebox{\plotpoint}}
\put(1050,511){\usebox{\plotpoint}}
\put(1052,510){\usebox{\plotpoint}}
\put(1053,509){\usebox{\plotpoint}}
\put(1054,508){\usebox{\plotpoint}}
\put(1055,507){\usebox{\plotpoint}}
\put(1056,506){\usebox{\plotpoint}}
\put(1058,505){\usebox{\plotpoint}}
\put(1059,504){\usebox{\plotpoint}}
\put(1060,503){\usebox{\plotpoint}}
\put(1061,502){\usebox{\plotpoint}}
\put(1062,501){\usebox{\plotpoint}}
\put(1064,500){\usebox{\plotpoint}}
\put(1065,499){\usebox{\plotpoint}}
\put(1066,498){\usebox{\plotpoint}}
\put(1067,497){\usebox{\plotpoint}}
\put(1068,496){\usebox{\plotpoint}}
\put(1070,495){\usebox{\plotpoint}}
\put(1071,494){\usebox{\plotpoint}}
\put(1072,493){\usebox{\plotpoint}}
\put(1073,492){\usebox{\plotpoint}}
\put(1074,491){\usebox{\plotpoint}}
\put(1076,490){\usebox{\plotpoint}}
\put(1077,489){\usebox{\plotpoint}}
\put(1078,488){\usebox{\plotpoint}}
\put(1079,487){\usebox{\plotpoint}}
\put(1080,486){\usebox{\plotpoint}}
\put(1082,485){\usebox{\plotpoint}}
\put(1083,484){\usebox{\plotpoint}}
\put(1085,483){\usebox{\plotpoint}}
\put(1086,482){\usebox{\plotpoint}}
\put(1087,481){\usebox{\plotpoint}}
\put(1089,480){\usebox{\plotpoint}}
\put(1090,479){\usebox{\plotpoint}}
\put(1091,478){\usebox{\plotpoint}}
\put(1093,477){\usebox{\plotpoint}}
\put(1094,476){\usebox{\plotpoint}}
\put(1095,475){\usebox{\plotpoint}}
\put(1096,474){\usebox{\plotpoint}}
\put(1097,473){\usebox{\plotpoint}}
\put(1098,472){\usebox{\plotpoint}}
\put(1100,471){\usebox{\plotpoint}}
\put(1101,470){\usebox{\plotpoint}}
\put(1102,469){\usebox{\plotpoint}}
\put(1103,468){\usebox{\plotpoint}}
\put(1104,467){\usebox{\plotpoint}}
\put(1106,466){\usebox{\plotpoint}}
\put(1107,465){\usebox{\plotpoint}}
\put(1108,464){\usebox{\plotpoint}}
\put(1109,463){\usebox{\plotpoint}}
\put(1110,462){\usebox{\plotpoint}}
\put(1111,461){\usebox{\plotpoint}}
\put(1112,460){\usebox{\plotpoint}}
\put(1113,459){\usebox{\plotpoint}}
\put(1114,458){\usebox{\plotpoint}}
\put(1115,457){\usebox{\plotpoint}}
\put(1117,456){\usebox{\plotpoint}}
\put(1118,455){\usebox{\plotpoint}}
\put(1120,454){\usebox{\plotpoint}}
\put(1121,453){\usebox{\plotpoint}}
\put(1122,452){\usebox{\plotpoint}}
\put(1124,451){\usebox{\plotpoint}}
\put(1125,450){\usebox{\plotpoint}}
\put(1126,449){\usebox{\plotpoint}}
\put(1128,448){\usebox{\plotpoint}}
\put(1129,447){\usebox{\plotpoint}}
\put(1130,446){\usebox{\plotpoint}}
\put(1132,445){\usebox{\plotpoint}}
\put(1133,444){\usebox{\plotpoint}}
\put(1134,443){\usebox{\plotpoint}}
\put(1136,442){\usebox{\plotpoint}}
\put(1137,441){\usebox{\plotpoint}}
\put(1138,440){\usebox{\plotpoint}}
\put(1140,439){\usebox{\plotpoint}}
\put(1141,438){\usebox{\plotpoint}}
\put(1142,437){\usebox{\plotpoint}}
\put(1144,436){\usebox{\plotpoint}}
\put(1145,435){\usebox{\plotpoint}}
\put(1146,434){\usebox{\plotpoint}}
\put(1148,433){\usebox{\plotpoint}}
\put(1149,432){\usebox{\plotpoint}}
\put(1150,431){\usebox{\plotpoint}}
\put(1152,430){\rule[-0.175pt]{0.361pt}{0.350pt}}
\put(1153,429){\rule[-0.175pt]{0.361pt}{0.350pt}}
\put(1155,428){\rule[-0.175pt]{0.361pt}{0.350pt}}
\put(1156,427){\rule[-0.175pt]{0.361pt}{0.350pt}}
\put(1158,426){\rule[-0.175pt]{0.361pt}{0.350pt}}
\put(1159,425){\rule[-0.175pt]{0.361pt}{0.350pt}}
\put(1161,424){\rule[-0.175pt]{0.361pt}{0.350pt}}
\put(1162,423){\rule[-0.175pt]{0.361pt}{0.350pt}}
\put(1164,422){\rule[-0.175pt]{0.361pt}{0.350pt}}
\put(1165,421){\rule[-0.175pt]{0.361pt}{0.350pt}}
\put(1167,420){\rule[-0.175pt]{0.361pt}{0.350pt}}
\put(1168,419){\rule[-0.175pt]{0.361pt}{0.350pt}}
\put(1170,418){\rule[-0.175pt]{0.361pt}{0.350pt}}
\put(1171,417){\rule[-0.175pt]{0.361pt}{0.350pt}}
\put(1173,416){\rule[-0.175pt]{0.361pt}{0.350pt}}
\put(1174,415){\rule[-0.175pt]{0.361pt}{0.350pt}}
\put(1176,414){\usebox{\plotpoint}}
\put(1177,413){\usebox{\plotpoint}}
\put(1178,412){\usebox{\plotpoint}}
\put(1180,411){\usebox{\plotpoint}}
\put(1181,410){\usebox{\plotpoint}}
\put(1182,409){\usebox{\plotpoint}}
\put(1184,408){\usebox{\plotpoint}}
\put(1185,407){\usebox{\plotpoint}}
\put(1187,406){\rule[-0.175pt]{0.361pt}{0.350pt}}
\put(1188,405){\rule[-0.175pt]{0.361pt}{0.350pt}}
\put(1190,404){\rule[-0.175pt]{0.361pt}{0.350pt}}
\put(1191,403){\rule[-0.175pt]{0.361pt}{0.350pt}}
\put(1193,402){\rule[-0.175pt]{0.361pt}{0.350pt}}
\put(1194,401){\rule[-0.175pt]{0.361pt}{0.350pt}}
\put(1196,400){\rule[-0.175pt]{0.361pt}{0.350pt}}
\put(1197,399){\rule[-0.175pt]{0.361pt}{0.350pt}}
\put(1199,398){\rule[-0.175pt]{0.413pt}{0.350pt}}
\put(1200,397){\rule[-0.175pt]{0.413pt}{0.350pt}}
\put(1202,396){\rule[-0.175pt]{0.413pt}{0.350pt}}
\put(1204,395){\rule[-0.175pt]{0.413pt}{0.350pt}}
\put(1205,394){\rule[-0.175pt]{0.413pt}{0.350pt}}
\put(1207,393){\rule[-0.175pt]{0.413pt}{0.350pt}}
\put(1209,392){\rule[-0.175pt]{0.413pt}{0.350pt}}
\put(1210,391){\rule[-0.175pt]{0.482pt}{0.350pt}}
\put(1213,390){\rule[-0.175pt]{0.482pt}{0.350pt}}
\put(1215,389){\rule[-0.175pt]{0.482pt}{0.350pt}}
\put(1217,388){\rule[-0.175pt]{0.482pt}{0.350pt}}
\put(1219,387){\rule[-0.175pt]{0.482pt}{0.350pt}}
\put(1221,386){\rule[-0.175pt]{0.482pt}{0.350pt}}
\put(1223,385){\rule[-0.175pt]{0.413pt}{0.350pt}}
\put(1224,384){\rule[-0.175pt]{0.413pt}{0.350pt}}
\put(1226,383){\rule[-0.175pt]{0.413pt}{0.350pt}}
\put(1228,382){\rule[-0.175pt]{0.413pt}{0.350pt}}
\put(1229,381){\rule[-0.175pt]{0.413pt}{0.350pt}}
\put(1231,380){\rule[-0.175pt]{0.413pt}{0.350pt}}
\put(1233,379){\rule[-0.175pt]{0.413pt}{0.350pt}}
\put(1234,378){\rule[-0.175pt]{0.482pt}{0.350pt}}
\put(1237,377){\rule[-0.175pt]{0.482pt}{0.350pt}}
\put(1239,376){\rule[-0.175pt]{0.482pt}{0.350pt}}
\put(1241,375){\rule[-0.175pt]{0.482pt}{0.350pt}}
\put(1243,374){\rule[-0.175pt]{0.482pt}{0.350pt}}
\put(1245,373){\rule[-0.175pt]{0.482pt}{0.350pt}}
\put(1247,372){\rule[-0.175pt]{0.530pt}{0.350pt}}
\put(1249,371){\rule[-0.175pt]{0.530pt}{0.350pt}}
\put(1251,370){\rule[-0.175pt]{0.530pt}{0.350pt}}
\put(1253,369){\rule[-0.175pt]{0.530pt}{0.350pt}}
\put(1255,368){\rule[-0.175pt]{0.530pt}{0.350pt}}
\put(1257,367){\rule[-0.175pt]{0.482pt}{0.350pt}}
\put(1260,366){\rule[-0.175pt]{0.482pt}{0.350pt}}
\put(1262,365){\rule[-0.175pt]{0.482pt}{0.350pt}}
\put(1264,364){\rule[-0.175pt]{0.482pt}{0.350pt}}
\put(1266,363){\rule[-0.175pt]{0.482pt}{0.350pt}}
\put(1268,362){\rule[-0.175pt]{0.482pt}{0.350pt}}
\put(1270,361){\rule[-0.175pt]{0.723pt}{0.350pt}}
\put(1273,360){\rule[-0.175pt]{0.723pt}{0.350pt}}
\put(1276,359){\rule[-0.175pt]{0.723pt}{0.350pt}}
\put(1279,358){\rule[-0.175pt]{0.723pt}{0.350pt}}
\put(1282,357){\rule[-0.175pt]{0.578pt}{0.350pt}}
\put(1284,356){\rule[-0.175pt]{0.578pt}{0.350pt}}
\put(1286,355){\rule[-0.175pt]{0.578pt}{0.350pt}}
\put(1289,354){\rule[-0.175pt]{0.578pt}{0.350pt}}
\put(1291,353){\rule[-0.175pt]{0.578pt}{0.350pt}}
\put(1294,352){\rule[-0.175pt]{0.723pt}{0.350pt}}
\put(1297,351){\rule[-0.175pt]{0.723pt}{0.350pt}}
\put(1300,350){\rule[-0.175pt]{0.723pt}{0.350pt}}
\put(1303,349){\rule[-0.175pt]{0.723pt}{0.350pt}}
\put(1306,348){\rule[-0.175pt]{0.723pt}{0.350pt}}
\put(1309,347){\rule[-0.175pt]{0.723pt}{0.350pt}}
\put(1312,346){\rule[-0.175pt]{0.723pt}{0.350pt}}
\put(1315,345){\rule[-0.175pt]{0.723pt}{0.350pt}}
\put(1318,344){\rule[-0.175pt]{0.883pt}{0.350pt}}
\put(1321,343){\rule[-0.175pt]{0.883pt}{0.350pt}}
\put(1325,342){\rule[-0.175pt]{0.883pt}{0.350pt}}
\put(1328,341){\rule[-0.175pt]{0.964pt}{0.350pt}}
\put(1333,340){\rule[-0.175pt]{0.964pt}{0.350pt}}
\put(1337,339){\rule[-0.175pt]{0.964pt}{0.350pt}}
\put(1341,338){\rule[-0.175pt]{1.445pt}{0.350pt}}
\put(1347,337){\rule[-0.175pt]{1.445pt}{0.350pt}}
\put(1353,336){\rule[-0.175pt]{0.964pt}{0.350pt}}
\put(1357,335){\rule[-0.175pt]{0.964pt}{0.350pt}}
\put(1361,334){\rule[-0.175pt]{0.964pt}{0.350pt}}
\put(1365,333){\rule[-0.175pt]{1.445pt}{0.350pt}}
\put(1371,332){\rule[-0.175pt]{1.445pt}{0.350pt}}
\put(1377,331){\rule[-0.175pt]{2.891pt}{0.350pt}}
\put(1389,330){\rule[-0.175pt]{2.650pt}{0.350pt}}
\put(1400,329){\rule[-0.175pt]{2.891pt}{0.350pt}}
\put(1412,328){\rule[-0.175pt]{2.891pt}{0.350pt}}
\put(1424,327){\rule[-0.175pt]{2.891pt}{0.350pt}}
\end{picture}
\caption{\sl Probability Distribution $Q(\theta^{\prime\prime})$}
\label{fig:pdq}
\end{figure}
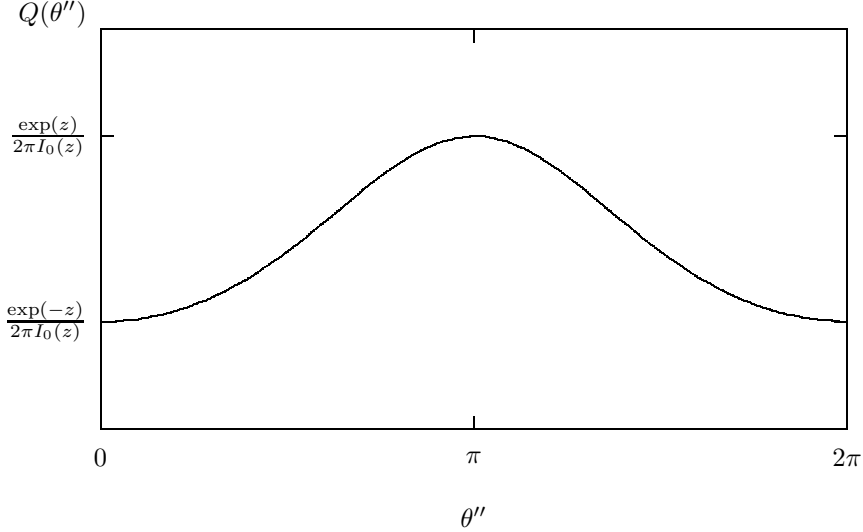

Similarly, we sample the outgoing dipolar orientations from
Eq.~(\ref{eq:nn7}).  Obtaining the outgoing angles from this, however,
is not straightforward; we need to sample this function but can not do
it exactly as in the case of the outgoing occupation numbers. The
probability distribution for each dipolar angle
$\theta_{i}^{\prime\prime}$, for $i= 1, \cdots, n$, is given by
\[
Q(\theta^{\prime\prime}) =
    \frac{\exp\left[-z \cos\theta^{\prime\prime}\right]}{2\pi
I_0(z)},
\]
where $z = \beta \nisp A_i(s)$ and $\theta^{\prime\prime} =
\theta_i^\prime-\phi_i(s)$. This function is that shown in
Fig.~(\ref{fig:pdq}).
For low $z$ we use a rejection algorithm to sample from this function.
This is done by choosing a random point, ($\theta$, $y$), uniformly
distributed in the rectangle of width $2\pi$ and height $e^z / (2\pi
I_0(z))$. If the chosen point is below the curve
$Q(\theta^{\prime\prime})$ then it is accepted and the first coordinate
of the point becomes the sampled $\theta^{\prime\prime}$; if the point
lies above the curve it is rejected and the procedure is repeated. One
problem with this method is that the efficiency of the required
algorithm goes as the ratio of the area under the curve to the area of
the rectangle, so that it becomes inefficient for large $z$.  It is,
however, reliable up to at least $z=6$.  For larger values of $z$ we
need another technique.  We observe that when $z$ is large the
distribution $Q(\theta^{\prime\prime})$ is very peaked near
$\theta^{\prime\prime} = \pi$ and can hence be approximated as follows:
\bgeas
Q(\theta^{\prime\prime}) & = &  \frac{\exp\left[-z
\cos\theta^{\prime\prime}\right]}{2\pi I_0(z)} \\
 & = & \frac{\exp\left[z
\cos\left(\theta^{\prime\prime} - \pi\right)\right]}{2\pi I_0(z)}\\
 & \approx &
 \frac{\exp\left[z\left(1 - \frac{1}{2}\left(\theta^{\prime\prime} -
\pi\right)^2 \right)\right]}{\frac{2\pi}{\sqrt{2\pi z}}\exp(z)}
\eeas
where we have used a Taylor expansion about $(\theta^{\prime\prime} -
\pi)$ and the asymptotic expansion of $I_0(z)$.  This gives us
\[
Q(\theta^{\prime\prime}) = \sqrt{\frac{z}{2\pi}}\exp\left[-\frac{z}{2}
\left(\theta^{\prime\prime} - \pi\right)^2\right].
\]
We observe that this is a normalised Gaussian in
$\theta^{\prime\prime}$, centered at $\pi$ and of width $\sqrt{1/z}$,
and consequently it is straightforward to sample $\theta^{\prime\prime}$
{}from this using the $2D$ Box-Mueller algorithm.

\section*{References}
\noindent Adler, C., d'Humi\`{e}res, D. \& Rothman, D.H. 1994 {\it J. Phys. I
(France)} {\bf 4}, 29-46.

\noindent Boghosian, B.M.,
Margolus, N.H. \& Yepez, J. ``Finite-Integer Lattice Gases,'' in preparation.

\noindent Cazabat, A.M., Langevin, D.,
Meunier, J. \& Pouchelon, A.
1982 {\it Adv. Colloid Interface Sci.} {\bf 16}, 175-199.

\noindent Chan, C.K. \& Liang, N.Y. 1990
{\it Europhys. Lett.} {\bf 13}, 495-500.

\noindent Coveney, P.V. \& Wattis, J.D. ``Analysis of a generalised
Becker-D\"{o}ring model of self-reproducing micelles,'' 1995 preprint.

\noindent Dawson, K.A. 1986 {\it Physical Review A.} {\bf 35}, 1766-1773.

\noindent Frisch, U., Hasslacher, B.
\& Pomeau, Y. 1986 {\it Phys. Rev. Lett} {\bf 56}, 1505-1508.

\noindent Gelbart, W.M., Roux, D. \&
Ben-Shaul, A. (eds) 1993 {\it Modern Ideas and Problems in Amphiphilic
Science.} Berlin: Springer.

\noindent Gompper, G. \& Schick, M.
1989 {\it Chem. Phys. Lett.} {\bf 163}, 475-479.

\noindent Gompper, G. \& Schick, M.
1995 {\it Phase Transitions and Critical Phenomena} {\bf 16}, to be published.

\noindent Gunstensen, A.K. \& Rothman, D.H. 1991a {\it Physica D.} {\bf 47},
47-52.

\noindent Gunstensen, A.K. \& Rothman, D.H. 1991b {\it Physica D.} {\bf 47},
53-63.

\noindent Gunstensen, A.K., Rothman, D.H., Zaleski, S. \& Zanetti,
G. 1991 {\it Phys. Rev. A} {\bf 43}, 4320-4327.

\noindent Gunstensen, A.K. 1992 {\it Ph.D. Thesis} M.I.T.

\noindent d'Humi\`{e}res, D. \& Lallemand, P. 1987 {\it Complex Systems}
{\bf 1}, 633-647.

\noindent Kahlweit, M., Strey, R., Haase, D., Kuneida, H. \&  Schmeling, T.
1987 {\it J. Colloid Interface Sci.}
{\bf 118}, 436.

\noindent Kawakatsu, T. \& Kawasaki, K. 1990 {\it Physica A} {\bf 167},
690-735.

\noindent Kawakatsu, T., Kawasaki, K., Furusaka, M., Okabayashi,
H. \& Kanaya, T. 1993 {\it J. Chem. Phys.} {\bf 99}, 8200-8217.

\noindent Matsen, M.W. \& Sullivan, D.E. 1990 {\it Phys. Rev. A} {\bf 41},
2021-2030.

\noindent Roland, C. \& Grant, M. 1989 {\it Physical Review B.} {\bf 39},
11971-11981.

\noindent Rothman, D.H. \& Keller, J.M. 1988 {\it J. Stat. Phys.} {\bf 52},
1119-1127.

\noindent Rothman, D.H. \& Zaleski, S.
1994 {\it Reviews of Modern Physics} {\bf 66}, 1417-1475.

\noindent Schick, M. \& Shih, W-H. 1987 {\it Phys. Rev. Lett.} {\bf 59},
1205-1208.

\noindent Stockfish, T.P. \& Wheeler, J.C. 1988, {\it J. Chem. Phys.}
{\bf 92}, 3292-3301.

\noindent Widom, B. 1984 {\it J. Chem. Phys.} {\bf 88}, 6508-6514.

\noindent Widom, B. 1986 {\it J. Chem. Phys.} {\bf 84}, 6943-6954.

\noindent Widom, B. \& Dawson, K.A. 1986 {\it Physica A.} {\bf 140}, 26-34.

\noindent Wolfram, S. 1986 {\it J. Stat. Phys.} {\bf 45}, 471-526.


\end{document}